\documentclass[pdflatex,sn-mathphys-num]{sn-jnl}

\usepackage{graphicx}%
\usepackage{multirow}%
\usepackage{amsmath,amssymb,amsfonts}%
\usepackage{amsthm}%
\usepackage{mathrsfs}%
\usepackage[title]{appendix}%
\usepackage{xcolor}%
\usepackage{textcomp}%
\usepackage{manyfoot}%
\usepackage{booktabs}%
\usepackage{algorithm}%
\usepackage{algorithmicx}%
\usepackage{algpseudocode}%
\usepackage{listings}%
\usepackage{lipsum}
\usepackage{color}
\usepackage{todonotes}

\definecolor{mytextcolor}{HTML}{2D6EBE}
\newcounter{suppfigure}
\newcounter{exdatafigure}
\makeatletter
\newcommand{\suppcaption}[1]{
  \refstepcounter{suppfigure}
  \begin{flushleft}
  {
    \small
    \addcontentsline{lof}{figure}{Supplementary Figure \thesuppfigure: #1}%
    \par
    \medskip
    \noindent\textbf{Supplementary Figure \thesuppfigure.} #1
    \par
  }
  \end{flushleft}
}
\makeatother

\makeatletter
\newcommand{\exdatacaption}[1]{
  \refstepcounter{exdatafigure}
  \begin{flushleft}
  {
    \small
    \addcontentsline{lof}{figure}{Extended Data Figure \theexdatafigure: #1}%
    \par
    \medskip
    \noindent\textbf{Extended Data Figure \theexdatafigure.} #1
    \par
  }
  \end{flushleft}
}
\makeatother

\DeclareMathOperator{\atanh}{atanh}

\begin{document}

\title[Article Title]{A neural network for modeling human concept formation, understanding and communication}


\author[1,2,3]{\fnm{Liangxuan} \sur{Guo}}
\equalcont{These authors contributed equally to this work.}

\author[4]{\fnm{Haoyang} \sur{Chen}}
\equalcont{These authors contributed equally to this work.}

\author*[1,3]{\fnm{Yang} \sur{Chen}}\email{yang.chen@ia.ac.cn}
\equalcont{These authors contributed equally to this work.}

\author*[4,5,6,7]{\fnm{Yanchao} \sur{Bi}}\email{ybi@pku.edu.cn}

\author*[1,2,3]{\fnm{Shan} \sur{Yu}}\email{shan.yu@nlpr.ia.ac.cn}

\affil[1]{\orgdiv{Laboratory of Brain Atlas and Brain-inspired Intelligence}, \orgname{Institute of Automation, Chinese Academy of Sciences}, \orgaddress{\state{Beijing 100190}, \country{China}}}

\affil[2]{\orgdiv{School of Future Technology}, \orgname{University of Chinese Academy of Sciences}, \orgaddress{\state{Beijing 100049}, \country{China}}}

\affil[3]{\orgdiv{State Key Laboratory of Brain Cognition and Brain-inspired Intelligence Technology}, \orgname{Institute of Automation, Chinese Academy of Sciences}, \orgaddress{\state{Beijing 100190}, \country{China}}}

\affil[4]{\orgdiv{School of Psychological and Cognitive Sciences \& Beijing Key Laboratory of Behavior and Mental Health}, \orgname{Peking University}, \orgaddress{\state{Beijing 100871}, \country{China}}}

\affil[5]{\orgdiv{IDG/McGovern Institute for Brain Research}, \orgname{Peking University}, \orgaddress{\state{Beijing 100871}, \country{China}}}

\affil[6]{\orgdiv{Institute for Artificial Intelligence}, \orgname{Peking University}, \orgaddress{\state{Beijing 100871}, \country{China}}}

\affil[7]{\orgdiv{Key Laboratory of Machine Perception (Ministry of Education)}, \orgname{Peking University}, \orgaddress{\state{Beijing 100871}, \country{China}}}


\abstract{A remarkable capability of the human brain is to form more abstract conceptual representations from sensorimotor experiences and flexibly apply them independent of direct sensory inputs. However, the computational mechanism underlying this ability remains poorly understood. Here, we present a dual-module neural network framework, the CATS Net, to bridge this gap. Our model consists of a concept-abstraction module that extracts low-dimensional conceptual representations, and a task-solving module that performs visual judgement tasks under the hierarchical gating control of the formed concepts. The system develops transferable semantic structure based on concept representations that enable cross-network knowledge transfer through conceptual communication. Model-brain fitting analyses reveal that these emergent concept spaces align with both neurocognitive semantic model and brain response structures in the human ventral occipitotemporal cortex, while the gating mechanisms mirror that in the semantic control brain network. This work establishes a unified computational framework that can offer mechanistic insights for understanding human conceptual cognition and engineering artificial systems with human-like conceptual intelligence.}



\maketitle

\section{Introduction}\label{sec1}

A unique feature of human language and thought, as pointed out by \textit{Ferdinand de Saussure} in 1916 \cite{de1916course}, is the ability to use “signifier” (for instance, symbolic reference) to communicate about “signified” referents that are physically absent. This capacity to decouple mental concepts from immediate sensory content allows humans to plan, simulate and represent information beyond the “here-and-now”. However, the computational framework that enables neural networks to form such concepts—initially dependent on sensorimotor stimuli but later independent of them—remains elusive. 

\begin{figure}[htbp]
\centering
\includegraphics[width=0.9\textwidth]{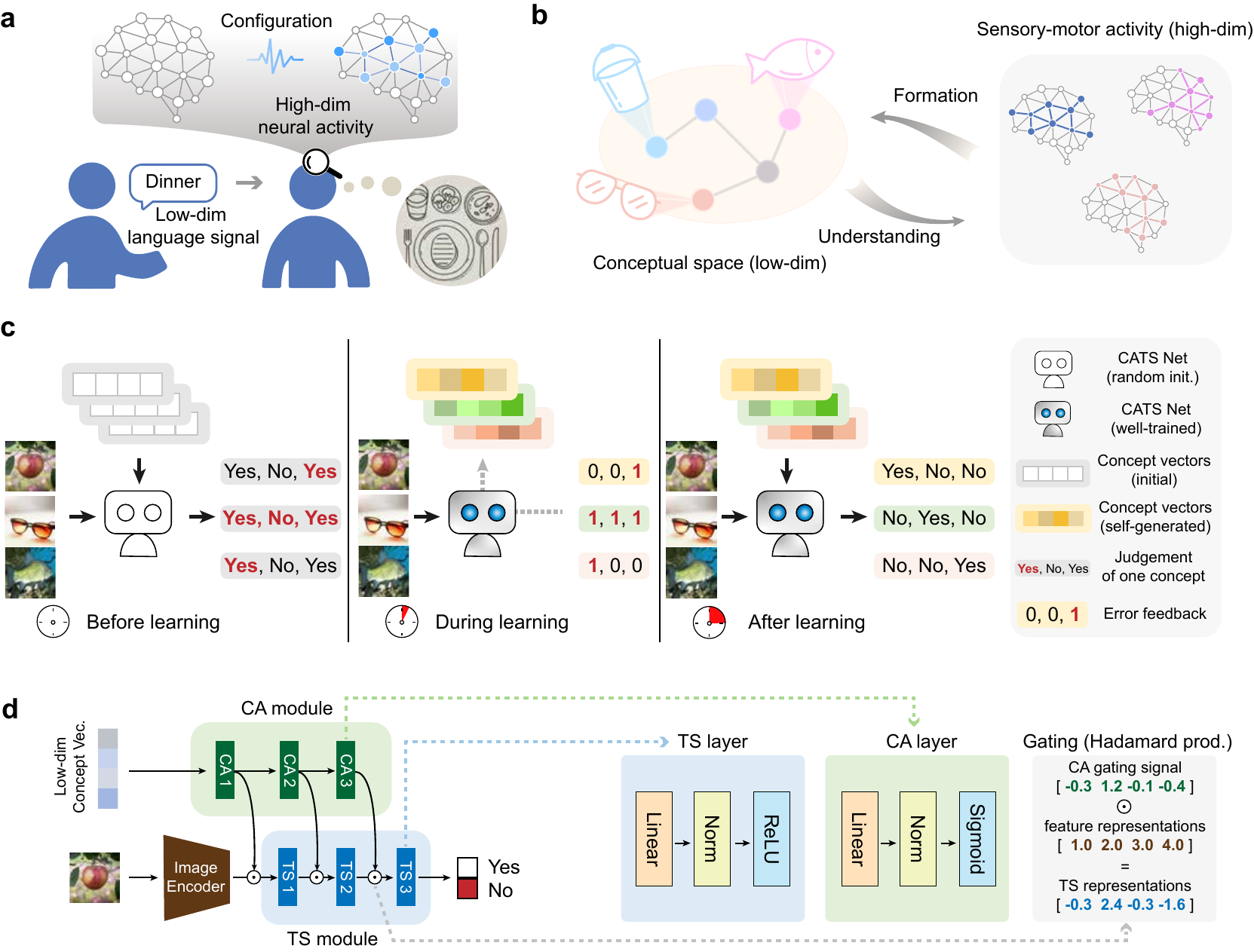}
\caption{\textbf{$\vert$ Motivation, experimental approach, and architecture of CATS net for concept decoupling and  formation. a,} The key characteristic of concepts is their decoupling from complex, high-dimensional sensory-motor information into lower-dimensional representations. For instance, the concept conveyed by a simple word like “dinner” can evoke neural population activity patterns associated with dining scenes, even without direct sensory stimulation. \textbf{b,}  A possible solution for concept formation is to compress sensory-motor neural circuits, independent of direct inputs, into low-dimensional representations. If these concepts can subsequently activate proper circuits to effectively accomplish the desired functions, it can be regarded as concept understanding. \textbf{c,} Illustration of our concept abstraction task approach. After training from random CATS Net parameter weights and initial concept vectors (all same or totally random), the system gets a set of well-trained parameter weights and well-trained concept vectors, which further support successfully making binary judgement for a given image under a given concept. \textbf{d,} Schematic illustration of the dual-module architecture in CATS net: the CA module receives low-dimensional conceptual inputs to generate controlling signal for TS module; The TS module performs “Yes/No” judgement according to sensory inputs and gating operation by CA module. All images were adopted from PublicDomainPictures and Free-Images under a Creative Commons license CC0.}
\label{fig1}
\end{figure}

In humans, concept processing comprises two coupled capacities: concept formation, where high-dimensional sensory-motor experiences are compressed into lower-dimensional representational spaces \cite{piantadosi_why_2024,grill-spector_functional_2014,binder_neurobiology_2011,ralph_neural_2017}, whose dimensionality typically ranges from 20 to several hundred \cite{mcrae_semantic_2005,devereux_centre_2014,binder_toward_2016,hebart_revealing_2020,haxby_common_2011}; and concept understanding, where these concepts are reactivated to reinstate sensorimotor states and flexibly combined \cite{pulvermuller_brain_2005,kiefer_conceptual_2012,huth_natural_2016,barsalou_grounded_2008,martin_grapesgrounding_2016,ralph_neural_2017}. For example, hearing "last night's dinner" would elicit rich event-related imagery (Figure~\ref{fig1}a), enabling communication of meanings through symbols. This bidirectional process is essential for concept processing in humans. A computational framework that simultaneously models both the concept formation and concept understanding remains a key challenge in artificial intelligence and neuroscience. 

Current approaches fall short of integrating these two functions. On one hand, deep neural networks like ResNet and Vision Transformers \cite{he_deep_2016,dosovitskiy_image_2021}, or classic CNNs with attention modules \cite{hu_squeeze-and-excitation_2018,woo_cbam_2018}, excel at learning representations, but entangle knowledge within millions of parameters, making it hard to decouple from the network, or directly transfer to another agent. On the other hand, Multimodal Large Language Models (MLLMs) \cite{radford_learning_2021,li_blip-2_2023,wu_deepseek-vl2_2024} rely on pre-existing language symbols rather than modeling de novo concept formation from sensorimotor experience.

Here, inspired by a previously proposed network capable of flexible context-dependent processing \cite{zeng_continual_2019}, we propose the CATS Net,a dual-module framework comprising concept-abstraction (CA) and sensorimotor task-solving (TS) modules (Figure~\ref{fig1}b). In this framework, concept formation is modeled as CA forming a low- dimensional input space of concept vector, while understanding is modeled via a gating mechanism where concept vectors dynamically reconfigure the TS module.

We demonstrated that CATS Net can derive novel concepts from a visual binary classification task, and transfer concept knowledge between CATS Nets. Importantly, concept representations formed in the CATS Net significantly correlate with human higher-order visual cortices, while the CA module aligns with the brain’s semantic control network \cite{jackson_neural_2021}, offering insights into the computational underpinnings of human conceptual processing.

\section{Results}\label{sec_result}
\subsection{Unified modeling of concept formation and understanding}\label{subsec_model_arch}

We introduced a concept abstraction task, where CATS Net generate a series of highly compressed concept vectors corresponding to particular visual category (Figure~\ref{fig1}c). Each learned vector functions as a functional classifier; for instance, an “apple” vector configures the network to judge whether an image input to the TS module belongs to apple category.

This is achieved via a hierarchical gating mechanism where the CA module transforms low-dimensional concept vectors into layer-wise control signals that modulate the TS module’s activity (Figure~\ref{fig1}d). And the TS module is a multi-layer perception (MLP) with a two-head classifier for binary (Yes/No) image judgment, to process features extracted by a pre-trained backbone. The framework’s design is agnostic to the backbone architecture, demonstrating robust generalization across different structures like ResNet50 \cite{he_deep_2016} and ViT-B/16 \cite{dosovitskiy_image_2021} (see Supplementary Figure 1a). 

The training process involves two phases: the network parameters learning phase, where the weights of CA module and TS module are trained together; and the concept abstraction phase, where the concept vectors are updated. These two phases were executed in a round-robin fashion until identification accuracy plateaued. This two- phase training strategy, along with the selection of a pretrained ResNet50 as the feature extractor, a concept size of 20, and 3-layer CA/TS modules, was validated as optimal through ablation studies (see Supplementary Figure 1a). Using this established configuration here and the rest of the current study, we trained 30 independently initialized models on the ImageNet-1k dataset \cite{deng_imagenet_2009}, which successfully generated a set of visual concept vectors for task solving. For all unseen images from 1000 categories tested on ImageNet-1k dataset, the learned concept vectors for each category achieved a judgement accuracy ranging from 0.86 to 1.00, well above the chance level of 0.5 (Figure~\ref{fig2}a).

Furthermore, through visualization, we observed that the models indeed focus on the part of the input image that corresponds to the concept. Using class activation mapping (CAM, \cite{selvaraju_grad-cam_2017}), with the same image input under different concept configurations, the network attends to different parts of the image (Figure~\ref{fig2}b). This shows that the network can adapt to different functions based on the conceptual input.

Importantly, our empirical comparisons indicate that the learnability of both the concept vectors and the CA/TS modules is equally critical. We compared our approach with three alternative methods for constructing a fixed concept space (Supplementary Figure 1b): using (1) frozen 20-dimensional random vectors, (2) frozen Word2Vec vectors projected to 20 dimensions, or (3) 1000-dimensional one-hot vectors.

The results revealed a crucial interplay between concept space learnability and network capacity. First, the trainable 20-dimension space significantly outperforms frozen and Word2Vec counterparts (random: mean difference = 0.0192, 95\% bootstrap CI [0.0185, 0.0200] with 5000 resamples, two-sided permutation test with 10,000 permutations, $p < 0.001$; Word2Vec: mean difference = 0.0279, 95\% bootstrap CI [0.0256, 0.0313] with 5000 resamples, two-sided permutation test with 10,000 permutations, $p < 0.001$). These results suggest that imposing a fixed concept space would force the network to compensate for arbitrary mappings, a constraint that becomes severe under limited capacity. For instance, reducing CA/TS modules from 3 layers to 1 layer, dropped accuracy with a frozen random space from 0.944 to 0.793, whereas enabling concept learnability restored accuracy to 0.954. 

Second, compared with the one-hot baseline, our approach was both more accurate and more scalable: a learnable 100-dimension concept space performed better (mean difference = 0.0043, 95\% bootstrap CI [0.0021, 0.0056], 5000 resamples, two-sided permutation test with 10,000 permutations, $p = 0.0079$) Furthermore, one-hot codes preclude any semantic structure, and scale linearly with the number of classes, requiring redefinition of the space when new concepts are added.

\subsection{Semantic Organization and Human Alignment of CATS Net}\label{subsec_space_prop}

\subsubsection{Functional Specificity of the Concept Space}

\begin{figure}[htbp]
\centering
\includegraphics[width=0.9\textwidth]{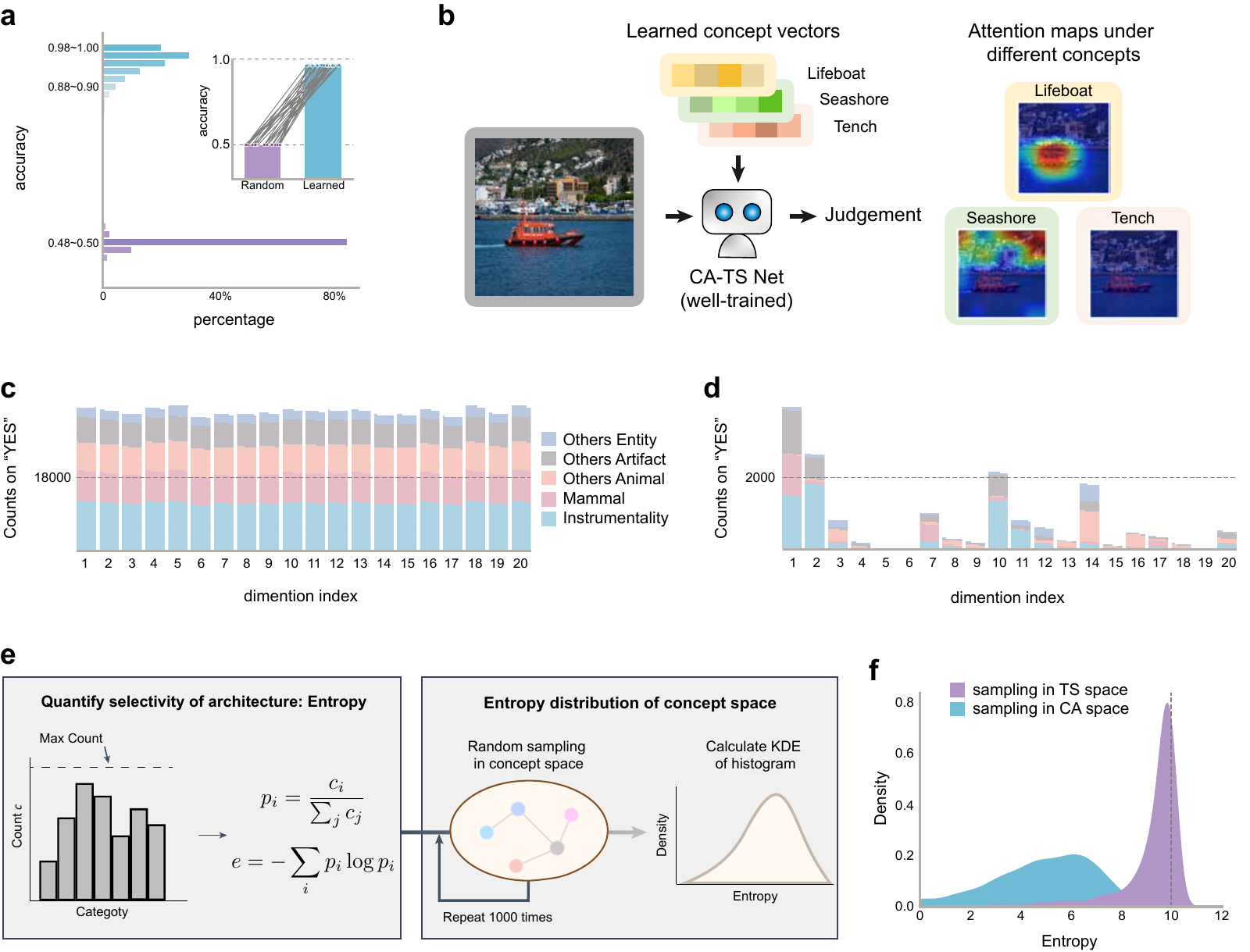}
\caption{\textbf{$\vert$ Model performance and conceptual space semantic structure analyses. a,} The performance of concept abstraction by CATS Net on ImageNet-1k dataset. The purple histogram depict the accuracy distribution of CATS Net for 1000 initial concept vectors before learning, while blue ones are after learning.  In the inset,  the purple and blue bar represents the average of mean accuracy across 30 models before and after training, and each pair point represents the corresponding mean accuracy of each category. \textbf{b,} Visualization of selective attentions on the same input modulated by different concept. \textbf{c\&d,} The functional specificity of the unit basis vectors on hyper-categories before (c) and after learning (d). The height of bar indicates the number of 'yes' response of the input basis vectors to these five hyper-categories. This result is a single case randomly chosen from 30 instances. \textbf{e,} Calculation pipeline of “functional entropy”, which quantitatively measures the functional specificity on the task. \textbf{f,} Probability density distribution of functional entropy in the trained concept space (blue) and task-solving parameter space (purple). All images were adopted from PublicDomainPictures under a Creative Commons license CC0.}
\label{fig2}
\end{figure}

To investigate the properties of the emergent 20-dimensional concept space, we probed its structure using the standard basis. Specifically, we employed the set of 20 canonical one-hot vectors, which referred to as the \textit{basis vectors} throughout the rest of this paper. Then we tested the functional specificity of the basis vectors. We categorized the 1,000 classes of ImageNet validation set into five hyper-categories based on WordNet, and then counted the “yes” response of the input basis vectors to these hyper-categories. Unlike the uniform response observed before training (Figure~\ref{fig2}c), trained basis vectors exhibited higher selectivity to specific hyper categories (Figure~\ref{fig2}d).

Taking a step further from the microscopic unit basis analysis to the macroscopic aspect, we introduce the ‘functional entropy’ to examine the overall functional specificity of the low-dimensional concept space. For a given concept vector, the functional entropy is computed over the number of ‘yes’ response counts of all 1000 classes (Figure~\ref{fig2}e, also see ‘Methods’). Low entropy corresponds to high functional specificity. We randomly sampled 1,000 points from the trained concept space; their entropy distribution was markedly lower than a random baseline (Figure~\ref{fig2}f), indicating a structured space with category-specific organization shaped by training.

\subsubsection{Representational Similarity to Human Semantic Models}

Next, we compared the CATS concept space with two complementary human semantic models: Binder65 (65 neurobiologically grounded dimensions \cite{binder_toward_2016}) and SPOSE49 (49 behavior-derived dimensions from THINGS similarity judgment \cite{hebart_revealing_2020}). Using representational similarity analysis (RSA) \cite{kriegeskorte_representational_2008} approach, we constructed RDMs over 332 shared concepts and correlated CATS RDMs with the semantic model RDMs at both the average-RDM level and across 30 independently trained CATS instances.

\begin{figure}[htbp]
\centering
\includegraphics[width=0.9\textwidth]{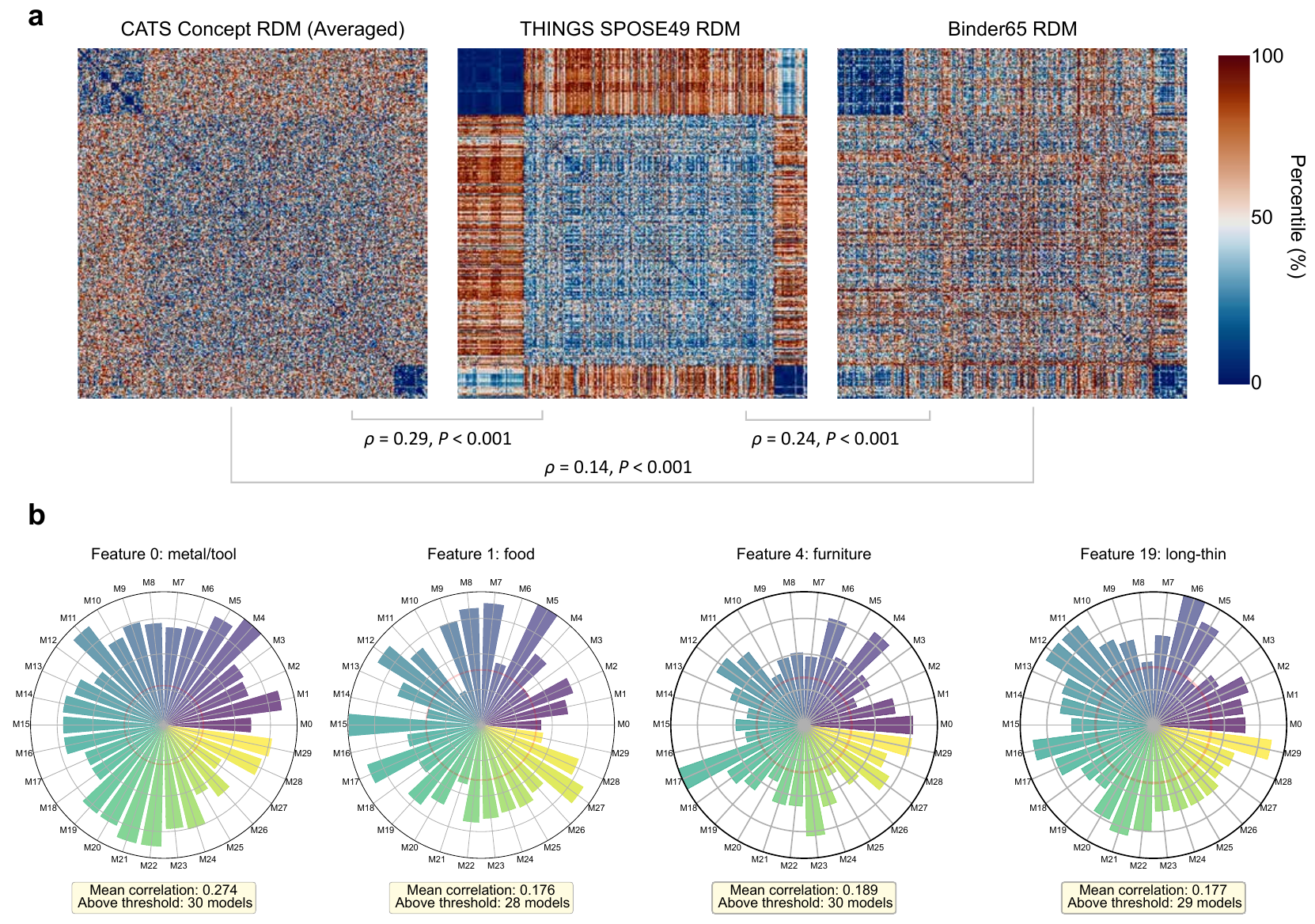}
\caption{\textbf{$\vert$ Alignment of CATS concept layer with human semantic models. a,} Representational dissimilarity matrices (RDMs) for the CATS concept layer, SPOSE49 \cite{hebart_revealing_2020}, and Binder65 \cite{binder_toward_2016} models, computed on 332 concepts overlapping between ImageNet-1k (\cite{deng_imagenet_2009}) and the THINGS dataset (\cite{hebart_things_2019}) using Pearson distance. The CATS RDM represents the average across 30 independently initialized instances. Warmer colors (red) indicate greater dissimilarity between concept pairs, while cooler colors (blue) indicate greater similarity. \textbf{b,} Correlations of best-match conceptual dimensions in each CATS Net instance with given four SPOSE49 exemplar dimensions. Each bar represents the maximum Pearson correlation between the 20 concept dimensions of a given CATS instance and a specific SPOSE49 dimension (labeled around the perimeter; dimension names from Hebart et al. \cite{hebart_revealing_2020}). The red circle denotes the significance threshold for correlation coefficients ($r = 0.107$, two-tailed $p < 0.05$, df = 330).}
\label{fig3}
\end{figure}

As shown in Figure~\ref{fig3}a, the average CATS Concept RDM showed significant correlations with Binder65 RDM (Spearman’s $\rho = 0.14$, Mantel $p < 0.001$) and SPOSE49 RDM (Spearman’s $\rho = 0.29$, Mantel $p < 0.001$). This correspondence was consistent across instances (Binder65, $t(29) = 18.28$, $p < 0.001$, Cohen’s $d = 3.39$, 95\% CI [0.03, 0.04]; SPOSE49, $t(29) = 25.43$, $p < 0.001$, Cohen’s $d = 4.72$, 95\% CI [0.06, 0.07]), indicating that CATS, despite being trained solely on visual categorization, generated a concept space similar to human conceptual organization. Notably, this similarity seems to be able to reflect CATS’s ability to capture abstract dimensions, as further evidenced by the significant correspondence with nonvisual dimensions of Binder65 (for instance, spatial, temporal, emotional; see Extended Data Figure~\ref{exdata_fig1}). 

\subsubsection{Semantic Interpretability}

To further explore the interpretability of our CATS Net concept space, we tried to provide semantic labels for our concept space dimensions. We employed the SPOSE49 model as a reference framework, because it captures finer-grained, visually relevant features that have been validated to effectively explain human similarity judgment behaviors \cite{hebart_revealing_2020} and neural response patterns \cite{contier_distributed_2024}. 

Specifically, we adopted a best-match procedure for each SPOSE49 dimension. Within each CATS instance, we computed Pearson correlations between a given SPOSE49 dimension and all concept dimensions of CATS Net, retaining the maximum correlation as the “best-match” score for that instance. This yielded, for each SPOSE49 dimension, a distribution of 30 best-match correlations across CATS instances. Figure~\ref{fig3}b highlighted four SPOSE49 dimensions (metal/tool, food, furniture, and long/thin) for which almost all 30 CATS instances exceeded the nominal significance threshold ($p < 0.05$; $n = 334$ concepts), indicating robust convergence on similar semantic structure despite different random initialization. The complete results across all 49 dimensions were presented in Supplementary Figure 2.

To directly visualize the structure of formed concept space by the CATS Net, we conducted identical task configuration training on a smaller-scale CIFAR-100 dataset \cite{krizhevsky_learning_2009}. Specifically, we got a set of 100 concept vectors by performing concept abstraction task on CIFAR-100. Then, we performed hierarchical clustering to these vectors based on cosine distance to visualize the internal structure of the low-dimensional concept space (Figure~\ref{fig4}a). This analysis reveals a modular structure characterized by distinct semantic clusters. Notably, semantically close categories formed clusters, for instance, clusters of people, animals, trees, fruit, furniture, and automobiles. These concept vectors enabled the establishment of connections among concepts through diverse multidimensional relationships, including similarities in foreground shape (like snakes and worms), foreground color (like sweet pepper and sunflower), background (like mushrooms and snail), and co-occurrence (like palm tree, cloud, and sea; tulip and butterfly).

\subsection{Communication by aligning the concept spaces of different CATS Nets}
\begin{figure}[htbp]
\centering
\includegraphics[width=0.9\textwidth]{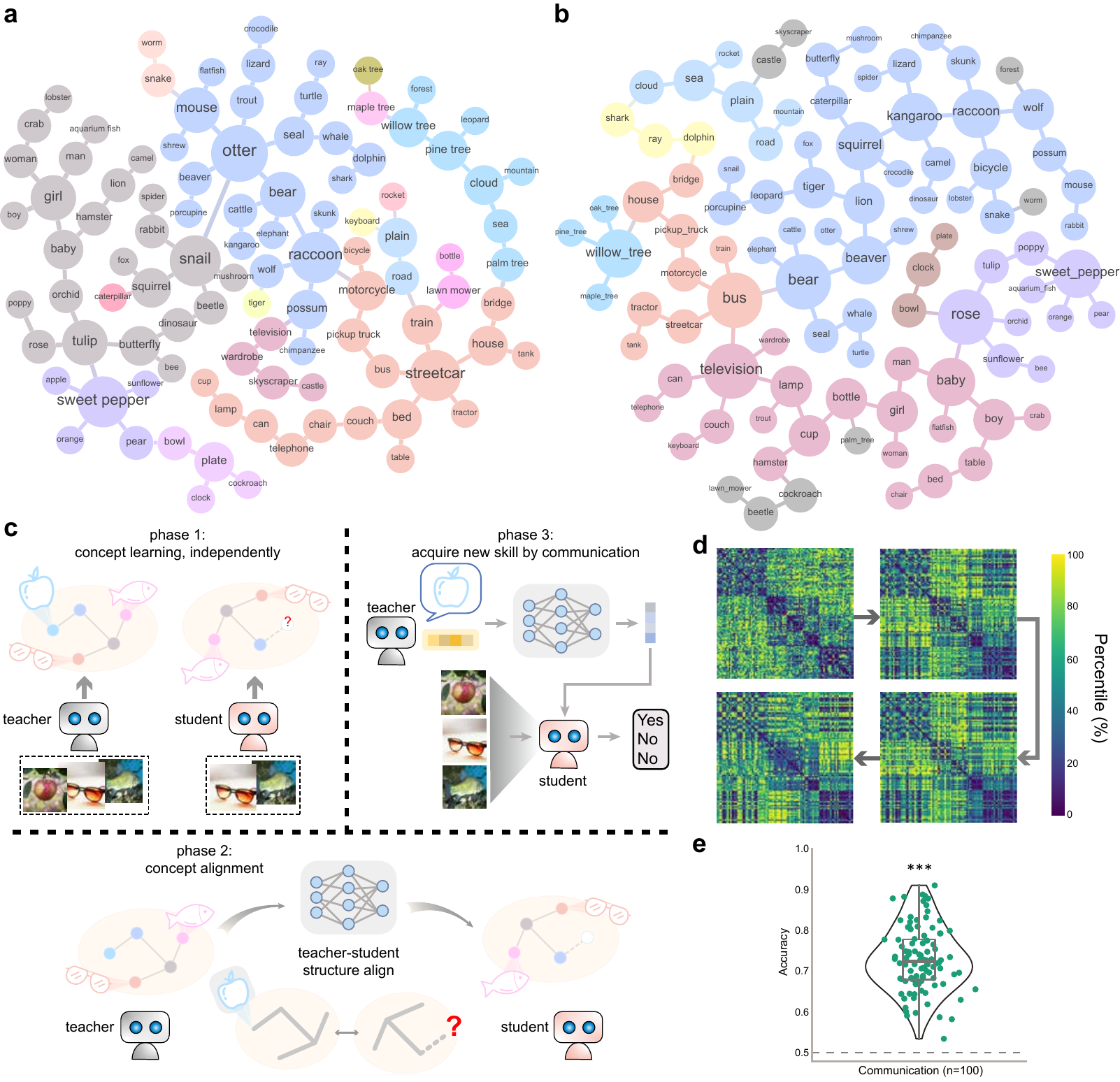}
\caption{\textbf{$\vert$ Knowledge transfer via communication between independently trained CATS Nets. a\&b,} Semantic maps of the concept space formed by the teacher (a) and the student networks (b). Colors represent clusters at a given hierarchical clustering threshold. Manual adjustments were then made to achieve the closest possible visual alignment between the teacher Net and student Net clusters in the visualization. \textbf{c,} Pipeline for knowledge acquisition via communication between the teach and student nets, consisting of three phases: independent concept abstracting, concept alignment and concept transmission. \textbf{d,} Layer-wise RDMs of the translation module (in order of arrows: input layer, $3^{rd}$ layer, $7^{th}$ layer, and output layer; see Supplementary Figure 3a for a complete view of all layers). \textbf{e,} Performance of transferred concepts on CIFAR-100 for student net through communication. Each dot represents the accuracy of an independent model instance (n=100 independent experimental units), where each was trained on a unique 99-category subset and evaluated on the corresponding held-out category. For all violin plots, individual dots represent independent model instances. The unit of analysis is a single model. Violin plots show the kernel density estimation of the data distribution. Overlaid box plots indicate the median (center line), interquartile range (IQR; 25th–75th percentiles), and min–max range (whiskers). Statistical significance for accuracy was determined by a one-tailed one-sample \textit{t}-test against chance level (0.5). $***$, $p < 0.001$; $**$, $p < 0.01$; $*$, $p < 0.05$. For all comparisons, the statistic values, degrees of freedom and exact P values are provided in the Source data. No technical replicates were used. Unless otherwise specified, the sample violin conventions are applied in all figures. All images were adopted from PublicDomainPictures and Free-Images under a Creative Commons license CC0.}
\label{fig4}
\end{figure}

Next, we define a “learning-by-communication” experiment to test if capabilities could be transferred solely via low-dimensional concepts. Using CIFAR-100 dataset, we ran 100 unique teacher–student pairs, each training the student network with one category held out. Each run comprised three phases: independent concept abstraction, concept alignment, and concept transmission (Figure~\ref{fig4}c).

\textbf{Phase 1: Independent Concept Abstraction.}
In the first phase,  an asymmetric training strategy was employed to create a knowledge gap: the teacher network learned all 100 categories, while the student network learned 99, withholding one distinct category (e.g., "apple") per pair for transfer testing. Despite independent initialization, the emerging concept spaces exhibited significant structural similarity. For instance, the modular organization of the teacher’s space (Fig. 4a) was mirrored in the student’s space (Figure~\ref{fig4}). Quantitative analysis using cosine-distance RDMs confirmed this alignment across all teacher–student pairs (Spearman's $\rho=0.35$, $t(99)=3.83$, one-tailed $p < 0.001$, Cohen's $d = 0.38$), indicating a shared internal structure for communication.

\textbf{Phase 2: Concept Space Alignment via a Translation Module.}
In the second phase, the two concept spaces were aligned with a translation module, that is, a neural network establishing a map from teacher concept space to student concept space, which was trained with expanded concept vectors (see Methods). To verify whether semantic details were preserved during translation, we analyzed the layer-wise internal representations of the translation module. Visualization of RDMs revealed that semantic clustering remained consistent across layers (Figure~\ref{fig4}d, Supplementary Figure 3a\&b). Through RDM correlation analysis across 100 teacher-student pairs (Supplementary Figure 3c), we found that the translation module systematically preserves semantic relationships while performing functional adaptation (Supplementary Figure 3d for statistical significance). Specifically, RDM correlations between input and successive layers showed a gradual decrease (from 0.93 to 0.29 at the output layer, all $p < 0.001$), indicating controlled information compression rather than arbitrary loss. 

\textbf{Phase 3: Concept Transmission and Evaluation.}
In the final phase, the teacher's novel concept vector (for instance, 'apple') was passed through the trained translation module, mapping it into the student's concept space. The student Net was then evaluated on its ability to perform yes/no judgments on input images using only this transferred concept vector. Across all 100 rounds of the experiment, the student networks demonstrated a remarkable ability to utilize the communicated concepts, performing consistently and significantly above the 0.5 chance level (mean accuracy (SD) $0.7292$ ($0.0781$), threshold $0.5$, $t(99)=29.33$, one-tailed $p < 0.001$, Cohen's $d=2.93$, 95\% CI [0.7137, 0.7448]).

These results validate the effectiveness of well-trained concepts in knowledge transfer. They imply that independently emerging concept spaces across separate networks share a common lexical-semantic structure, enabling the acquisition of new knowledge without updating the high-dimensional network parameters.

\subsection{Compatibility of CATS Net with Human Language- and Behavioral-Derived Concept Spaces} 

To test whether our CATS Net is able to directly utilize the concept space generated by humans, we evaluated its performance using both language-derived and direct human behavioral-derived concept spaces.

\begin{figure}[htbp]
\centering
\includegraphics[width=0.9\textwidth]{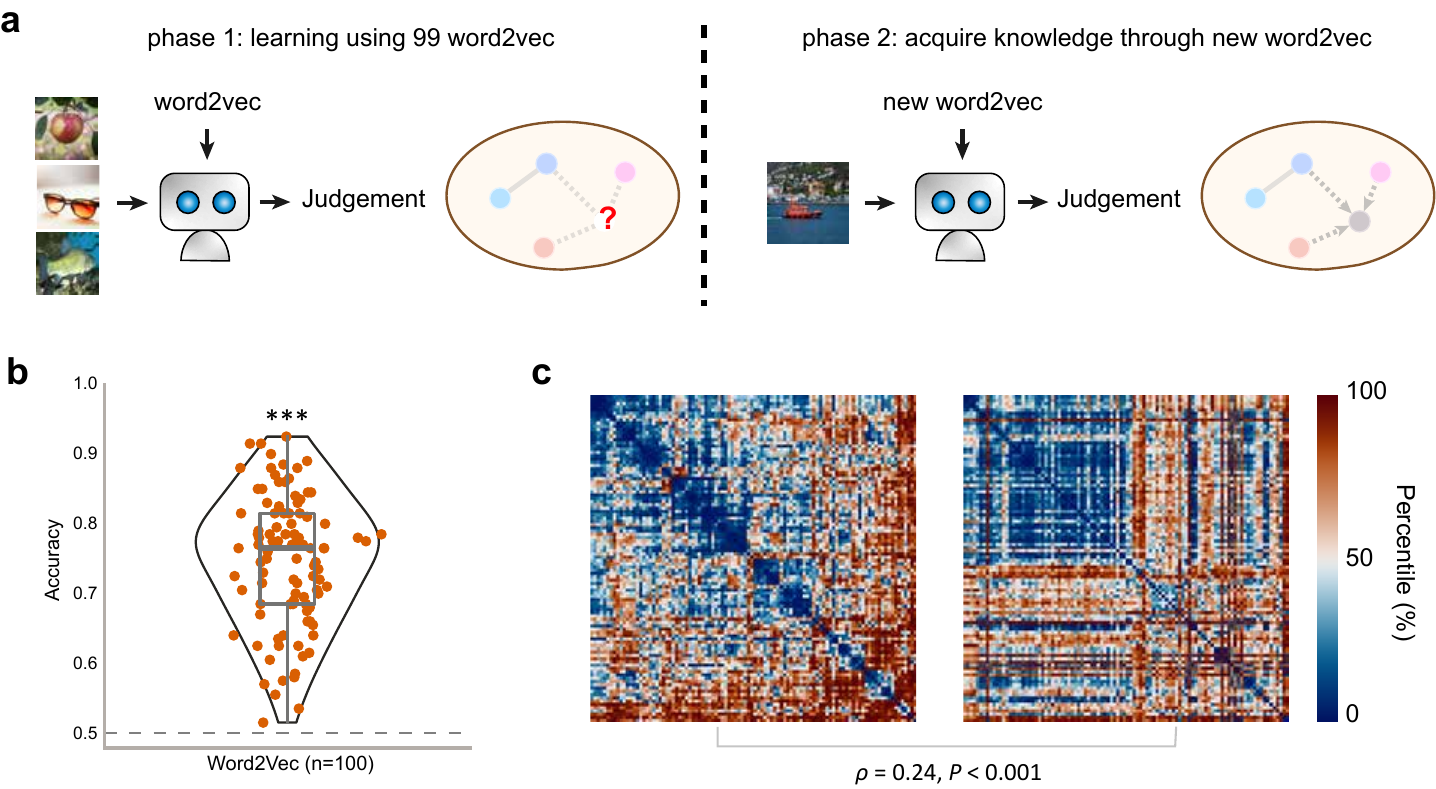}
\caption{\textbf{$\vert$ Concept acquisition in CATS Net using Word2Vec embeddings. a, }Pipeline for novel concept acquisition in CATS Net on CIFAR-100 using Word2Vec. In Phase 1, images from 99 categories and their name embeddings (as predefined concept vectors) are used to train a randomly initialized CATS Net by updating only network parameters. In Phase 2, the remaining category and its Word2Vec embedding (as a unseen concept vector) is used to evaluate the model’s understanding of the novel concept. \textbf{b,} Performance on unseen concepts under the leave-one-out approach described in (a). Each dot represents the accuracy of an independent model instance ($n = 100$ independent experimental units), where each was trained on a unique 99-category subset and evaluated on the corresponding held-out category. \textbf{c,} RDMs of learned concept vectors (left) versus Word2Vec vectors (right). All images were adopted from PublicDomainPictures and Free-Images under a Creative Commons license CC0.}
\label{fig5}
\end{figure}

\textbf{Word2Vec-Based Concept Space.} We first used the Word2Vec space as the low-dimensional concept space and evaluated the ability of these human language-derived vectors to configure the TS module in the CATS Net. To this end, we conducted a leave-one-out concept abstraction experiment, which is composed of two phases (Figure~\ref{fig5}a). Specifically, in the first phase, the system was trained with fixed Word2Vec from 99 concepts, and learnable network parameters (both CA and TS modules), as described above. Individual category names, represented by their corresponding word vectors, were projected into 20-dimensional space and fed into the CA module. In the second phase, the remaining category (also referred to as the conceptual inferred category) was evaluated by performing yes/no judgement given an unlearned word vector as the concept input. By travelling through all categories with leave-one-out approach, we can demonstrate that the accuracy of each conceptual inferred category is well beyond the chance level (mean accuracy (SD) $0.7474$ ($0.0934$), threshold $0.5$, $t(99)=26.51$, one-tailed $p < 0.001$, Cohen's $d = 2.65$, 95\% CI [0.7289, 0.7660]). Despite never encountering the images or category names before, the system successfully recognized majority of images (Figure~\ref{fig5}b).

Additionally, we assessed the similarity between the concept vectors generated by the CATS Net and Word2Vec's vector representations. Although the Word2Vec vectors are derived from word co-occurrence statistics in large text corpora, which is fundamentally different from our model, their RDMs still show a significant correlation with ours (Spearman's $\rho=0.24$, Mantel $p < 0.001$, 10,000 permutations; bootstrap 95\% CI [0.154, 0.366], 5,000 resamples; Figure~\ref{fig5}c). 

\textbf{Human Behavioral Data-Based Concept Space.} To further validate our architecture's compatibility with human behavior-derived concept spaces, we conducted experiments using SPOSE49 model. Using these human-generated concept vectors, we replicated the leave-one-out experimental approach described above. The results demonstrate that our CATS Net architecture achieves comparable performance when configured with human behavioral data (mean accuracy (SD) $0.6967$ ($0.1582$), threshold $0.5$, $t(99)=12.43$, one-tailed $p < 0.001$, Cohen's $d = 1.24$, 95\% CI [0.6653, 0.7281]), confirming that our dual-module framework can effectively exploit not only language-derived concept spaces but also genuine human perceptual and conceptual structures (Extended Data Figure~\ref{exdata_fig2}). This validation strengthens the claim that our architecture provides a general computational framework for concept formation and understanding that is compatible with authentic human cognitive processes.

These results collectively suggest that the framework proposed here can effectively exploit the information structure in human's concept space to solve new tasks, providing a common computational framework for concept formation and understanding across totally different systems.

\subsection{Mapping neural representations: visual and semantic control networks}

\begin{figure}[htbp]
\centering
\includegraphics[width=0.7\textwidth]{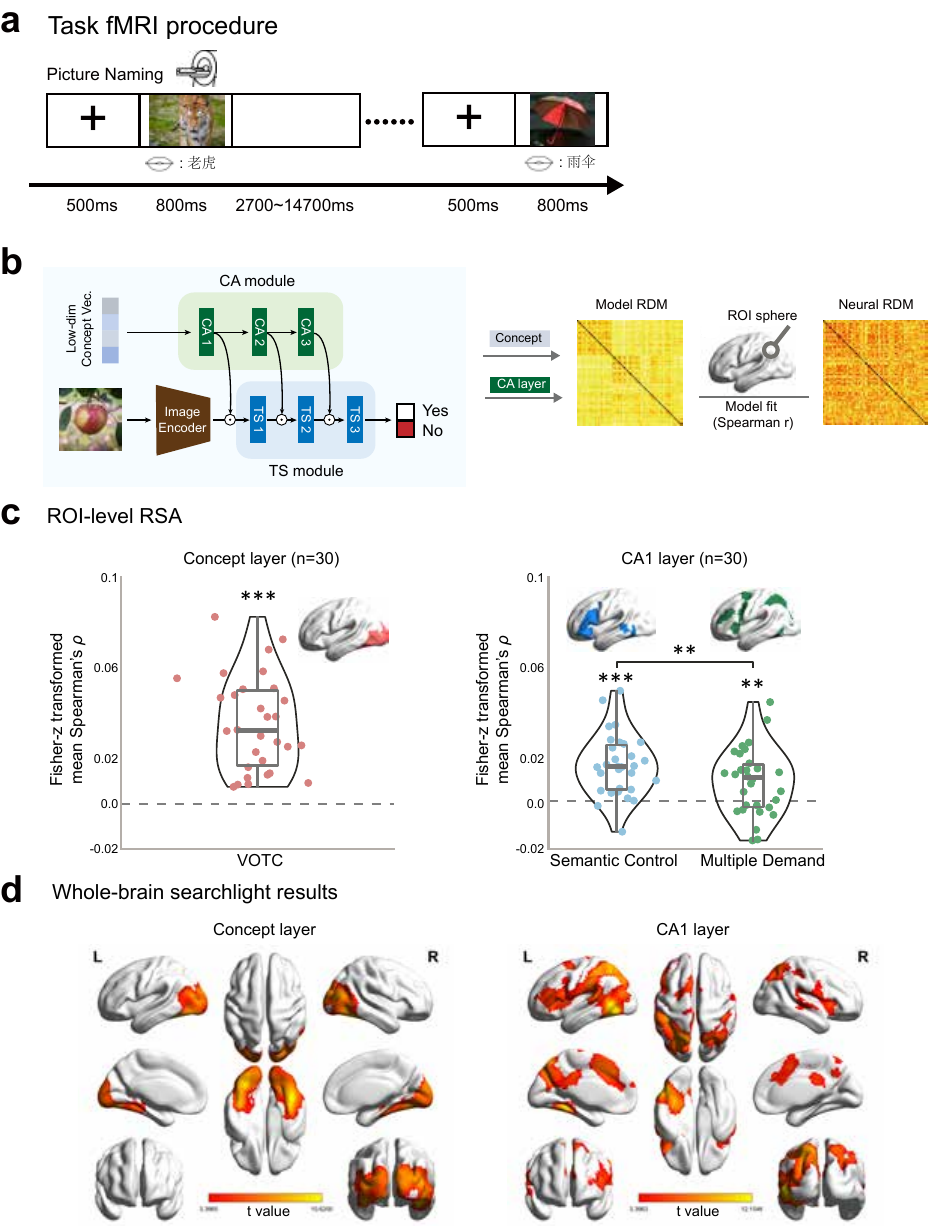}
\caption{\textbf{$\vert$ Representational similarity between CATS Net layers and human brain. a,} Task fMRI experimental design for object naming. The pictures used in the fMRI task were changed to pictures under a Creative Commons license CC BY 4.0. \textbf{b,} Representational Similarity Analysis pipeline. Neural RDMs were correlated with model RDMs to quantify correspondence. \textbf{c,d,} ROI-level (c) and whole-brain searchlight (d) RSA results. For both analyses, correlations were first averaged across 26 subjects for each independently trained model (n=30). \textbf{c,} Correspondence between the concept layers of CATS Nets and VOTC (left), and between the CA1 layer and semantic control\cite{jackson_neural_2021}/multi-demand networks\cite{fedorenko_broad_2013} (right). Each dot represents one model. For single-group comparisons, significance was determined by a one-tailed one-sample  $t$-test against zero. Between-group differences were assessed using paired \textit{t}-test. Asterisks indicate significance levels: $**$, $p < 0.01$; $***$, $p < 0.001$. \textbf{d,} Group-level \textit{t}-value maps for concept layer (left) and CA1 layer (right). Searchlight results were thresholded at voxel-level $p < 0.001$, one-tailed, and cluster-level family-wise error (FWE)-corrected $p < 0.05$. All images were adopted from Free-Images under a Creative Commons license CC0.}
\label{fig6}
\end{figure}

\begin{figure}[htbp]
\centering
\includegraphics[width=0.8\textwidth]{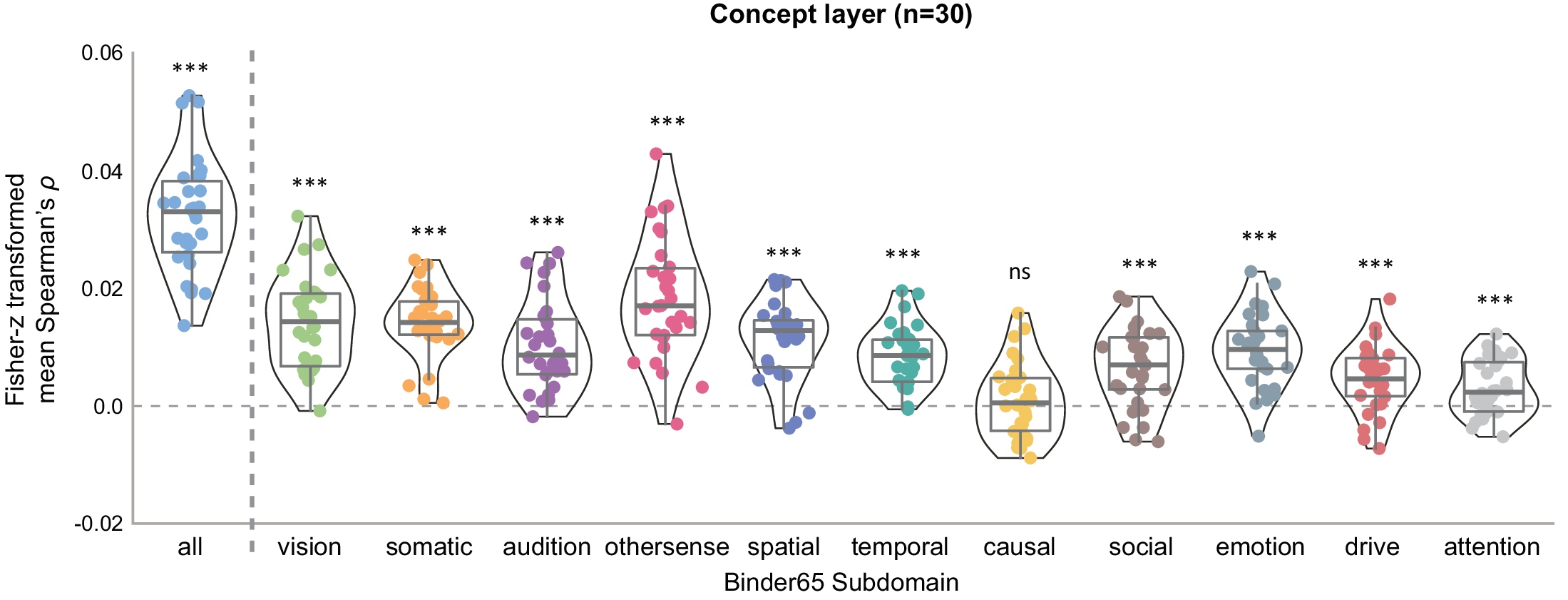}
\exdatacaption{\textbf{$\vert$ Representational similarity between CATS concept layer and Binder65 subdomains on ImageNet dataset.} This figure illustrates the RSA results between our CATS model's concept layer and the 11 subdomains of the Binder65 model. RDMs were generated for both the CATS concept layer and each Binder65 subdomain using WT95 stimulus dataset. The y-axis displays Fisher's \textit{z}-transformed Spearman’s rank correlation coefficients ($\rho$) between the respective RDMs. Individual data points represent correlation values from each independently trained model. Asterisks ($***$) above each subdomain indicate statistical significance ($p < 0.001$) from one-sample \textit{t}-tests conducted at the group level. The "ns" indicate $p > 0.05$.}
\label{exdata_fig1}
\end{figure}

\begin{figure}[htbp]
\centering
\includegraphics[width=0.9\textwidth]{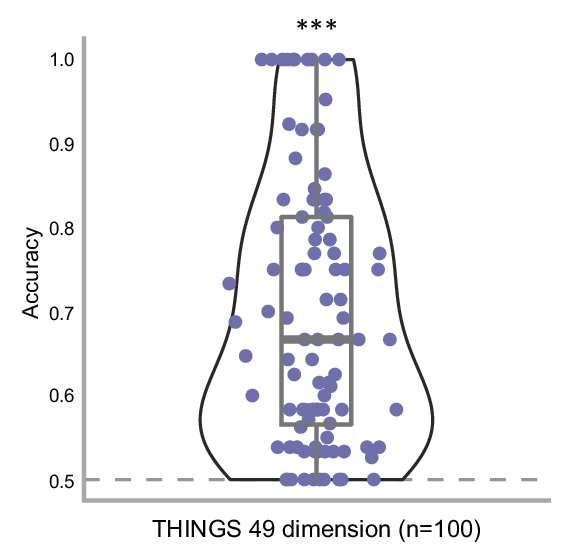}
\exdatacaption{\textbf{$\vert$ Performance on unseen concepts under the leave-one-out approach using THINGS and 49-dimension human-generated embedding vectors.} We randomly chose 100 categories from 1,854 object categories from THINGS as 100 independent runs of leave-one-out experiment. Each point represents the accuracy of one category chosen for leave-one-out experiment.}
\label{exdata_fig2}
\end{figure}

To assess the extent to which the concept spaces generated by our CATS Net models align with those in the human brain, we compared similarity patterns between the model concept spaces and the visual cortex activities during object perception task using RSA. These analyses utilized our previously published fMRI dataset \cite{fu_different_2022}, which contained human brain activity data to 95 objects that covers three common object domains (32 animals, 35 small manipulable artefacts, and 28 large non-manipulable artefacts). Participants ($N=26$ in the analyses) viewed images presented on the screen and performed an oral picture-naming task (Figure~\ref{fig6}a). Given that our models are trained solely on visual categorization tasks, we first conducted a region-of-interest (ROI) based RSA targeting the ventral occipitotemporal cortex (VOTC) for object perception \cite{ungerleider_what_1994}, defined by contrasting picture viewing versus rest (FDR $q < 0.05$; see Methods for details). We computed the partial Spearman’s rank correlation between the model’s concept layer and human VOTC representations, explicitly controlling for low-level visual features (that is, the ResNet sensory input layer).

As shown in Figure~\ref{fig6}c (left panel), the representational patterns of concept layers of CATS Net model showed highly significant correlation with human VOTC activity patterns(Fisher-\textit{z} mean (SE) $\rho = 0.04$ $(0.004)$, $t(29) = 9.27$, one-tailed $p < 0.001$, Cohen's $d = 1.70$). This indicates that the abstraction mechanism in CATS Net captures conceptual representations aligning with human neural coding significantly beyond canonical visual features.

\begin{figure}[htbp]
\centering
\includegraphics[width=0.9\textwidth]{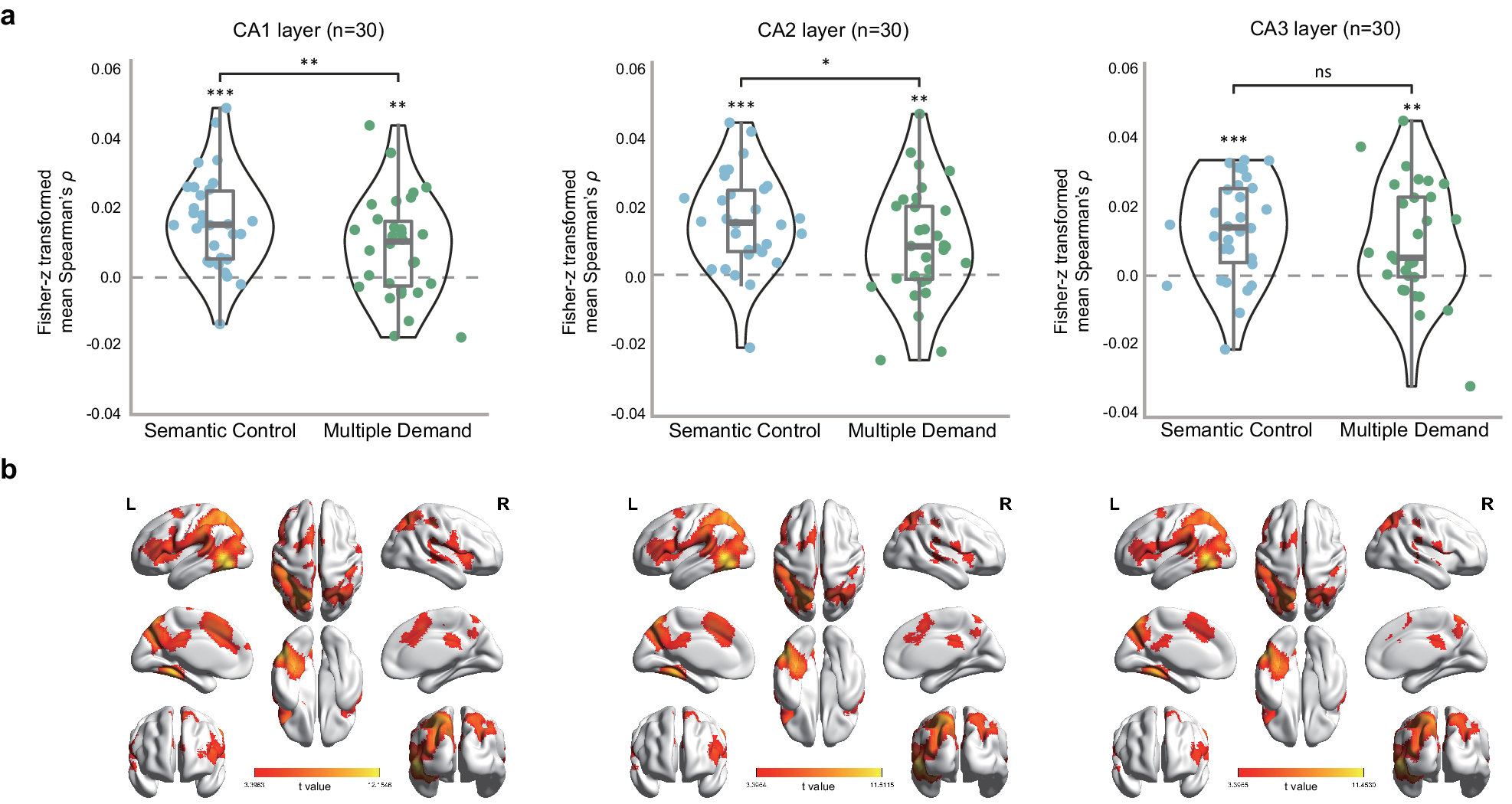}
\exdatacaption{\textbf{$\vert$ RSA model-group analysis results for the three layers of CA module across 30 independent model instances.} The top panel presents the results of ROI analysis, with semantic control network \cite{jackson_neural_2021} and domain-general multi-demand (domain-general control) network \cite{fedorenko_broad_2013} used as ROIs. For single-group comparisons, significance was determined by a one-tailed one-sample  $t$-test against zero. Between-group differences were assessed using paired \textit{t}-test. Asterisks indicate significance levels: $**$, $p < 0.01$; $***$, $p < 0.001$. The bottom panel displays the whole-brain searchlight RSA results, at the threshold of voxel-level $p < 0.001$, one-tailed, cluster-level family-wise error (FWE)-corrected $p < 0.05$, highlighting the spatial patterns of model-brain correspondence across the whole brain. Each of the three subplots corresponds to one layer (CA1, CA2, and CA3) of the models.}
\label{exdata_fig3}
\end{figure}

We next examined the CA module. This module dynamically gates feature representations, a function analogous to the human semantic control network, which has been assumed to selectively access and manipulate meaningful conceptual information in relevance to a particular context or task \cite{jackson_neural_2021}. Since the semantic control network modulates access to information rather than representing visual content itself, we assessed alignment without controlling for the sensory input layer. The first layer of the CA module (CA1) showed significant correspondence with the semantic control network (Fisher-\textit{z} mean (SE) $\rho = 0.02$ $(0.003)$, $t(29) = 6.44$, one-tailed $p < 0.001$; additional CA module layers also demonstrated significant correlations, see Extended Data Figure~\ref{exdata_fig3}). Crucially, this alignment showed functional specificity: while the CA1 layer also correlated with the domain general multiple demand (MD) network \cite{fedorenko_broad_2013} (Fisher-\textit{z} mean (SE) $\rho = 0.01$ $(0.003)$, $t(29) = 3.22$, one-tailed $p < 0.01$) (Fig~\ref{fig6}c right panel; for other CA layers, see Extended Data Figure~\ref{exdata_fig3}a), the alignment with the semantic control network was significantly stronger (paired \textit{t}-tests: Fisher-\textit{z} mean difference (SE) $= 0.01$ $(0.003)$, $t(29) = 2.89$, two-tailed $p < 0.01$). These findings collectively suggest that the CA module aligns specifically with the semantic-control processes.

Across ROIs, effect size was reliable across 30 independently initialized models (one-sample tests on Fisher-\textit{z} means) and should be interpreted relative to the noise ceilings (VOTC–concept $NC_z = 0.25$; Multiple-Demand–CA1 $NC_z = 0.27$; Semantic-Control–CA1 $NC_z = 0.24$; see Methods). For context, the VOTC correspondence of a widely used baseline model (ResNet-50) and SPOSE49 model were 0.007 and 0.056, respectively (mean Fisher-z transformed $\rho$, mean across 26 participants). 

\begin{figure}[htbp]
\centering
\includegraphics[width=0.9\textwidth]{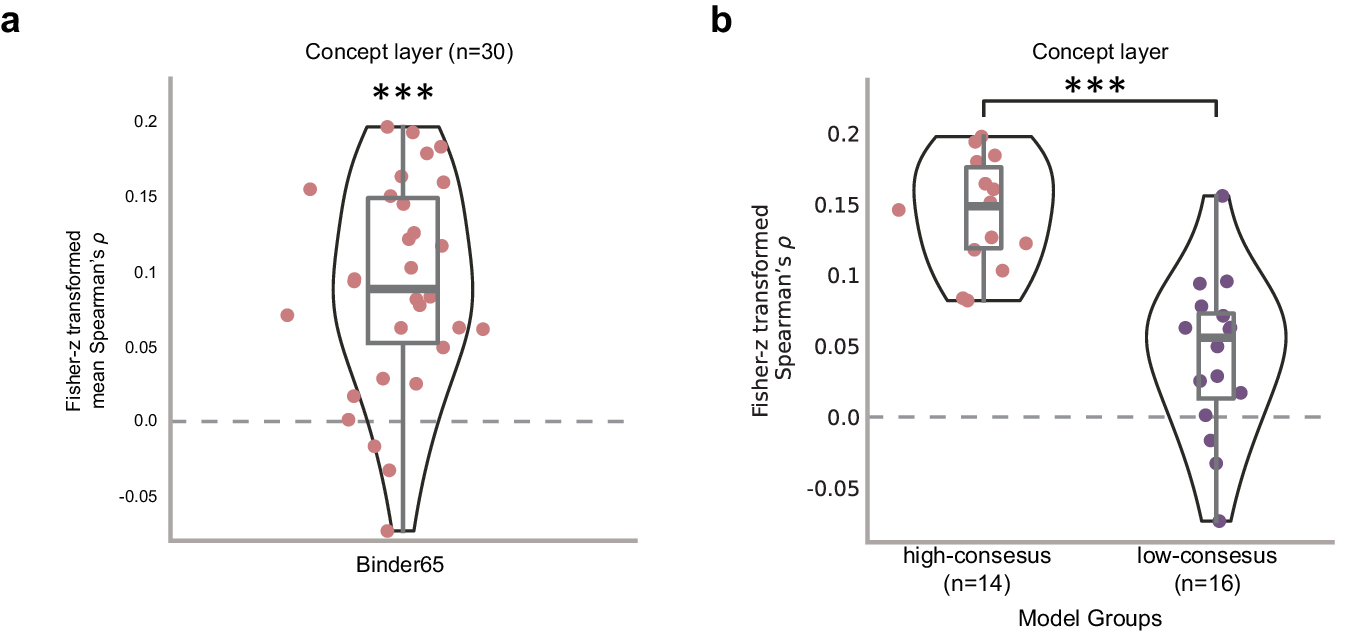}
\exdatacaption{\textbf{$\vert$ CATS Net-Binder65 correspondence using RSA.} Box plots showing the Spearman’s correlation coefficients (Fisher-\textit{z} transformed) between our CATS Nets and the Binder65 model based on RSA analysis, controlling for the sensory input layer. The left panel displays the group-level results for all 30 models, with each point representing an independently trained model instance. The right panel presents the results of a cluster analysis dividing the models into high-consensus (n=14) and low-consensus groups (n=16), with subsequent group-level analysis and between-group comparison. Each point in this panel also represents a model instance within the respective group. Statistical comparison between groups were performed using a two-tailed two-sample \textit{t}-test ($***$, $p < 0.001$).}
\label{exdata_fig4}
\end{figure}

Finally, we confirmed these ROI-based findings using whole-brain searchlight RSA (Fig~\ref{fig6}b) \cite{kriegeskorte_information-based_2006}. At the threshold of voxel-level $p < 0.001$, one-tailed, cluster-level family-wise error (FWE)-corrected $p < 0.05$, we found that the concept layer across all 30 CATS Nets showed significant correspondence with the bilateral VOTC (Figure~\ref{fig6}d left panel). In contrast, the CA1 layer corresponded to regions typically associated with the semantic control network, including the bilateral dorsomedial prefrontal cortex (bi-dmPFC), inferior parietal lobe (bi-IPL), left inferior frontal gyrus (l-IFG), lateral occipital complex (l-LOC) and posterior fusiform gyrus (l-pFG) (Figure~\ref{fig6}d right panel; for other CA layers, see Extended Data Figure~\ref{exdata_fig3}b). These findings aligned with our ROI results, confirming that the concept layer corresponds to neural representations in VOTC, while the CA module predominantly corresponds to activities in semantic control regions (for validation, see Extended Data Figure~\ref{exdata_fig3} and \ref{exdata_fig4}, Supplementary Figure 4).

\subsection{Emergent cross-model consensus increases alignment with human semantic systems}
Our previous analyses demonstrated that model-generated concept representations showed significant model-group level correlations with human visual cortex activity. Here we zoom into the individual model spaces to examine whether convergent patterns among independently trained CATS Nets might predict stronger alignment with human semantic systems.

\begin{figure}[htbp]
\centering
\includegraphics[width=0.9\textwidth]{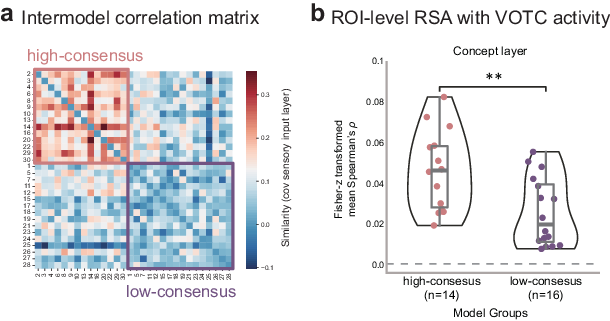}
\exdatacaption{\textbf{$\vert$ Clustering analysis and group-level RSA of 30 independently trained models.} Based on a two-class clustering approach, models were categorized into a high-consensus group (14/30) and a low-consensus group (16/30). Representational similarity analysis (RSA) was conducted to evaluate correspondence between each group's concept layers and human brain activity in the ventral occipitotemporal cortex (VOTC). \textbf{a,} Inter-model correlations of 30 independent trained models. Models are categorized into high-consensus (n=14) and low-consensus (n=16) groups based on representational similarity. \textbf{b,} RSA results for CATS Net's concept layer with VOTC activity. Each dot represents a single model's mean Spearman's correlation across 26 subjects. Results are plotted separately for high-consensus and low-consensus groups. Statistical comparison between groups were performed using a two-tailed two-sample \textit{t}-test ($**$, $p < 0.01$).}
\label{exdata_fig5}
\end{figure}

Biological neural systems often exhibit evolutionary convergence in their semantic coding strategies \cite{carandini_normalization_2012}, suggesting optimal solutions emerge under similar hardware constraints. To test whether this principle applies to our networks, we first performed cluster analysis on 30 independently trained CATS Nets. This analysis revealed a dominant organizational pattern emerging in 47\% of models (14/30; Extended Data Figure ~\ref{exdata_fig5} left panel). We defined these 14 models as our “high-consensus” group, hypothesizing that their shared representational structure might indicate greater biological plausibility compared to the remaining 16 models (the “low-consensus” group).

To evaluate this hypothesis, we conducted two complementary analyses examining the alignment between these model groups and two independent measures of human semantic representation. First, we assessed alignment with the Binder65 neurobiological semantic model. High-consensus models demonstrated significant superior correspondence with Binder65 (mean difference (SE) $= 0.11$ $(0.01)$, $t(29) = 6.98$, two-tailed $p < 0.001$; Extended Data Figure~\ref{exdata_fig4}), suggesting the emergent consensus structures capture neurobiologically-relevant semantic dimensions. Second, we tested whether this consensus advantage extended to alignment with actual brain activity. High-consensus models also showed stronger correspondence with VOTC activity patterns than low-consensus models (mean difference (SE) $= 0.02$ $(0.00)$, $t(29) = 3.05$, two-tailed $p < 0.01$; Extended Data Figure~\ref{exdata_fig5} right panel), while controlling for the pretrained sensory input layer RDM. These findings imply that both artificial and biological systems might be governed by equivalent optimization imperatives in how they organize semantic information. The spontaneous emergence of shared representational geometry across independently trained networks suggests that when subjected to comparable computational constraints, different systems—biological or artificial— tend to converge toward similar semantic coding solutions. This convergence may reflect fundamental organizational principles that efficiently support semantic processing across different types of intelligent systems.

\section{Discussion}

The CATS Net architecture offers a unified computational framework for linking raw sensory experience with symbolic thought. By integrating concept formation through compression and understanding through sensorimotor reinstatement, this dual-module system provides a computational account of how high-dimensional sensory inputs are mapped into low-dimensional, communicable conceptual spaces. This process moves beyond the limitations of purely language-derived representations by demonstrating that functionally useful conceptual structures can emerge directly from task-driven sensorimotor grounding \cite{harnad_symbol_1990}. Such a mechanism is particularly vital for capturing nuanced or domain-specific knowledge that is often difficult to articulate fully through natural language—a form of intuitive, embodied expertise that can now be computationally instantiated and utilized within artificial neural networks.

The implications of this grounded representational mechanism also suggest a different perspective on emergent communication. While many current approaches rely on end-to-end joint optimization via backpropagation \cite{foerster_learning_2016, jaques_social_2019,wang_emergence_2024}, CATS Net motivates a more modular and biologically plausible alternative: agents can develop internal conceptual structures independently and subsequently align them through a shared symbolic interface. This approach captures a foundational principle of human interaction, where low-dimentional symbols are used to reactivate rich, high-dimensional sensorimotor experiences in others. In principle, such a design may reduce the dependence on continual, large-scale joint re-optimization and facilitate incremental alignment in decentralized multi-agent settings, providing a plausible route towards scalable collective intelligence.

Beyond its implications for artificial intelligence, the framework offers a concrete hypothesis that connects to established neurocognitive accounts. The observed correspondence between CATS representations and brain activity patterns in human VOTC and semantic control network is consistent with theories proposing top-down modulation of feature representations during semantic processing \cite{ralph_neural_2017}. In this view, task-related feature gating—implemented here as a multiplicative, element-wise interaction—serves as a candidate mechanism through which relevant semantic features are amplified and irrelevant ones suppressed according to shifting task demands. While other forms of modulation exist, formalizing this multiplicative account helps translate descriptive theories into testable computational predictions about the neural basis of semantic flexibility.

However, the current scope of CATS Net is primarily constrained to concrete concepts with identifiable visual referents. Abstract concepts (for instance, "justice" or "freedom") pose additional challenges due to their lack of bounded physical referents and high inter-individual variability \cite{borghi_challenge_2017, barsalou_abstraction_2003}. Although our exploratory analyses suggest that CATS Net's concept layers capture representations related to abstract dimensions, the model was not explicitly optimized for the fuzzy boundaries and limited annotation reliability characteristic of non-perceptual categories \cite{martinezpandiani_hypericons_2023}. A key direction for future work is therefore to extend this architecture to multi-modal integration, examining how the joint constraints of vision, audition, and language can further sharpen conceptual boundaries and support more abstract representational spaces. More broadly, systems that combine conceptual compression with sensorimotor reinstatement may offer a practical step towards AI that represents—and communicates about—the world in a more human-like, grounded manner.

\section{Methods}
\subsection{Hierarchical gating of CATS Net}
In concept abstraction task on image dataset, we use a 20-dimension real-valued vector to present each category. This dimension was selected from a tested range of {10, 20, 100}, as it offered the optimal trade-off between compression efficiency and representational capacity (Supplementary Figure 1a). The compactness of this vector space, compared to the high-dimensional parameter space of the neural network, reflects the highly compressed nature of the concepts. The model's pipeline begins with a pretrained ResNet50 backbone \cite{he_deep_2016}, chosen over ViT \cite{dosovitskiy_image_2021} for its computational efficiency after observing similar performance from both (we use official V1 weights from \url{https://pytorch.org/} for both backbones). The extracted 2,048-dimension features are then fed into the TS module. This module is a 3-layer perceptron ([2,048-100-100-2]) with batch normalization and ReLU activation. The 3-layer architecture was adopted for its demonstrated robustness, as our tests with 1, 3, and 5 layers all yielded comparable performance (see Supplementary Figure 1a). To match this structure, the CA module is also a 3-layer perceptron ([20-2,048-100-100]), which takes the 20-dimension concept vector as input and uses the Sigmoid function $\sigma(x) = \frac{1}{1 + e^{-x}}$ to generate controlling signals between 0 and 1. The output layer of the TS module consists of two neurons for "Yes" (0, 1) and "No" (1, 0) classification, optimized using a cross-entropy loss. 

Consider a MLP of $L+1$ layers, indexed by $l=0,\cdots,L$ with $l = 0$ and $l = L$ being the input and output layers. Let $W_l^{TS}$ be the connection weight between the $(l-1)^{th}$ layer and $l^{th}$ layer in the TS module, while $W_l^{CA}$ for the CA module. $\mathbf{x}_{l-1}$ denotes the input of the connections $W_l^{TS}$, while $\mathbf{c}_{l-1}$ denotes the input of the connections $W_l^{CA}$, respectively. $\mathbf{o}_{l}$ denotes the output of the connections $W_l^{TS}$, while $\mathbf{g}_{l}$ denotes the output of the connections $W_l^{CA}$, respectively. It is clear that the dimension of feature extractor output is the same with the $\mathbf{x}_{0}$ and the dimension of concept vector is with $\mathbf{c}_{0}$. For the CA module, after applying normalization and activation to the $\mathbf{g}_{l}$, it will be the input $\mathbf{c}_{l}$ for weight $W_{l+1}^{CA}$.

The CA module does not need to directly modulate the feature extractor, because it is possible to utilize gating signals in a much easier way to modulate the processing of the feature extractor. The dimensions of $\mathbf{x}_{l-1} \in \mathbb{R}^{d}$ and $\mathbf{g}_{l} \in \mathbb{R}^{d}$ are consistent from $l=1$ to $L$. For $l=1$ to $L$, let $\mathbf{z}_{l-1} = \mathbf{x}_{l-1} \odot \mathbf{g}_{l}, \mathbf{z}_{l-1} \in \mathbb{R}^{d}$, we replace the $\mathbf{x}_{l-1}$ by $\mathbf{z}_{l-1}$ and set it as input for weight $W_{l}^{TS}$. Operator $\odot$ is the Hadamard product (also known as the element-wise product), is a binary operation that takes in two matrices of the same dimensions and returns a matrix of the multiplied corresponding elements. In our case, the hierarchical gating take place at [2,048-100-100] layers between CA and TS modules.

Under this network structure, even if the same stimulus $\mathbf{x}$ is provided to the TS module, the CATS Net will conduct hierarchical gating operations based on the different concept vector given to the CA module. Let $H(\mathbf{x}, \mathbf{c})$ be the CATS Net, so $H(\mathbf{x}, \mathbf{c}) = G(T(\mathbf{x}), C(\mathbf{c}))$ where $T(\cdot)$ is the TS module, $C(\cdot)$ is the CA module and $G$ is the hierarchical gating between CA module and TS module. When such gating only takes effect at a certain layer, it is equivalent to scaling the data of the current layer. And when it acts on multiple layers along with activation functions, provided that the input stimulus $\mathbf{x}$ remains unchanged, there exists a variation of $T'$ s.t. $H(\mathbf{x}, \mathbf{c}) = T'(\mathbf{c})$. That is to say, it is equivalent to realizing several different TS parameters with distinct concept vectors.

\subsection{Concept abstraction task data and training}

The input to the CATS Net include concept vector and natural images while the output is a 2-dimension vector indicating “Yes” or “No”. So the original image-label doublet vision dataset will be convert to the image-concept-label triplet one. Take ImageNet-1k dataset used in the current work as an illustration, we randomly sampled 1,000 points in a 20-dimension real vector space and assigned them to each category, which was fed to the CA module as the initial for the abstracted concept. Then for half of images in the whole dataset, we assign the corresponding concept vector to them, along with the “Yes” label. While for another half, we randomly assign a non-corresponding concept vector with label of “No”, as negative samples for training stability. 

Two training phases first begin with the network learning phase: the concept vector inputs to the CA module were fixed, while all network parameters, including those in both CA and TS modules, were updated by gradient back-propagation using a binary supervising signal (Yes/No), to indicate whether the image belongs to the target concept category or not. In the following concept learning phase, only concept vectors were modified by the back-propagated gradients, with all network parameters fixed. The two phases in the training process were carried out alternatively in an epoch-by-epoch manner. It provides better interpretability of concept learning dynamics, which implies the learning of concept space can be independent from the learning of network parameters. We also validated this approach against end-to-end joint training and found comparable performance (Supplementary Figure 1a), confirming that the concept formation process is robust to training methodology. The training was terminated after 5 epochs to ensure accuracy reaching the plateau. Uniform distributed noise ranging from -0.1 to 0.1 was injected into each element of the concept vectors, in both the network learning phase and concept learning phase. We found that it effectively increased the system’s robustness for distinguishing various categories. No noise was added to the concept vectors in the testing. In all experiments, the length of concept vectors was set to 20, that is, they contained 20 real elements.

\subsection{Visualization of configured CATS Net by CAM}
We made a little modification for traditional Grad-CAM \cite{selvaraju_grad-cam_2017}, to directly show the importance of each neuron in the last layer of the pretrained feature extractor, after being gated by the signals from CA module. In order to obtain the class-discriminative localization map $L_{Grad-CAM}^{\mathbf{c}} \in \mathbb{R}^{u \times v}$ of width $u$ and height $v$ for any concept $\mathbf{c}$, we first compute the gradient of the “Yes” score $y^\mathbf{c}$ (the value of “Yes” neuron before the softmax, given concept input), with respect to feature maps $A^k$ of a convolutional layer, that is, $\frac{\partial y^\mathbf{c}}{\partial A^k}$. These gradients flowing back are global average-pooled to obtain the neuron importance weights $\alpha_k^\mathbf{c}$,
\[
\alpha_k^\mathbf{c} = g_{1, k} \frac{1}{Z} \sum_i \sum_j \frac{\partial y^\mathbf{c}}{\partial A^k_{ij}}
\]
where $i$ and $j$ are the index for width $u$ and height $v$, that is, the pixel in each 2D convolution kernel, and $Z$ is the total number of pixel in this kernel. $k$ stands for the index of convolution kernel, it is straight forward that the number of convolution kernels is the same as the dimension of the gating vector $\mathbf{g}_1$. The $g_{1, k}$ is the $k^{th}$ element of $\mathbf{g}_1$, that is, gating signals from the output of $W_1^{CA}$. 

This weight $\alpha_k^\mathbf{c}$ represents a partial linearization of the deep network downstream from $A$, and captures the “importance” of feature map $k$ for a target concept $c$. We perform a weighted combination of forward activation maps, and follow it by a ReLU to obtain,
\[
L_{Grad-CAM}^{\mathbf{c}} = ReLU(\sum_k \alpha_k^\mathbf{c} A^k)
\] 
Finally, $L_{Grad-CAM}^{\mathbf{c}}$ is linearly scaled to the size of the input image so as to obtain the activation map shown in Fig~\ref{fig2}b.

\subsection{Hyper-category functional specificity of the basis vector of concept space}
We assigned a hyper-category label to each category in ImageNet-1k, using wordnet from nltk library. Specifically, since each class label in ImageNet has its corresponding synset ID in WordNet \cite{fellbaum_wordnet_1998}, we first obtain all the hyper synsets of each class in WordNet to form a WordNet synset-chain for that class. Subsequently, we examine the synsets one by one from the top of the synset-chain downwards to check whether they correspond to the four preset hyper-categories such as mammals and artifacts. If none of them can be matched, then the hyper-category label of “others entity” will be assigned to that class. The synset tokens for 4 hyper-category were “mammal.n.01”, “animal.n.01”, “instrumentality.n.03” and “artifact.n.01”. Thus, the 5 hyper-categories were “mammal”, “others mammal”, “instrumentality”, “others artifact” and “others entity”.

For a well-trained CATS Net, given the one-hot vectors ranging from dimension 1 to 20, we calculated the number of images with a “Yes” response over the evaluation set of ImageNet-1k (50,000 images), for each hyper-category.

\subsection{Functional entropy}
For each concept vector in the concept space, we define the functional entropy as 
\[
\displaystyle e=-\sum_{i} p_i \log p_i
\]
where 
\[
\displaystyle p_i = \frac{c_i}{\sum_{j}c_j}
\]
The $c_i$ stands for the number of “Yes” response to $i^{th}$ class across the whole classes in the dataset, while the $p_i$ is the normalized probability prepared for entropy calculation. A higher value of functional entropy also implies that the current input concept vector has a relatively even selectivity for each category. In other words, this vector cannot represent the concept of a specific category in the dataset. On the contrary, a lower entropy indicates that the concept vector is more inclined to respond “Yes” to certain specific categories while answering “No” for the majority of the remaining categories. So the distribution of the functional entropy reflect the overall attribution over the whole concept space.

\subsection{Hierarchical clustering analysis of CIFAR-100 concept set}
We used hierarchical clustering  (Matlab function dendrogram) to group concept vectors generated by CATS Net, based on cosine distance between vectors and unweighted average linkage between clusters. Specifically, two concepts, each from one of two distinct but connected branches or leaves in the dendrogram with the closest distance, were connected by one edge. We traversed all pairs of connected branches and leaves, linking all pairs of concept nodes to meet the requirement of the closest distance. The visualization of the semantic network was generated by Gephi \cite{bastian_gephi_2009}.

\subsection{Technical implementation of communication experiment: Leave-one-out training and concept vector expansion}
This section describes the technical procedures underlying the communication experiment presented in Figure~\ref{fig4}. The leave-one-out training and concept vector expansion serve two critical functions: (1) creating knowledge asymmetry between teacher and student networks, and (2) generating sufficient training data for the translation module that enables concept transfer.

For the leave-one-out training, the student CATS Net was trained on dataset $D_{99}$ containing images and labels from 99 categories, while one category $D_1$ was withheld to create the knowledge gap that communication aims to bridge. The teacher Net was trained on the complete dataset including $D_1$. 

Subsequently, to generate expanded concept vectors for training the translation module, concept vector expansion for the withheld category $D_1$ was performed through concept manipulation only, that is, only through the concept abstraction phase without retraining the network parameters. To utilize the concept obtained by so far to identify $D_1$ as much as possible, we introduced a repelling loss $L_{rep}$ for learning new concept, which was defined as
\[
L_{rep}(C, C_{old}, \tau) = \sum_{C_i \in C_{old}} \exp (-|C_i - C|^2/\tau)
\]
where $C_i \in C_{old}$ were the concepts of categories belonging to $D_{99}$ and $C$ the concept of the remaining category in $D_1$. To test the system’s capability of few shot learning, only two images belonging to $D_1$ and one image from each of the $99$ learned categories belonging to $D_{99}$ were utilized in concept abstraction. The concept assigned to the category in $D_1$ were randomly initialized and trained to minimize the following loss function
\[
L = L_{CE}(x_{new},y|C) + \alpha L_{CE}(x_{old},\bar{y}|C) + \beta L_{rep}(C, C_{old}, \tau)
\]
where $L_{CE}$ denotes the cross-entropy loss, $x_{new}$ the image sampled from the new category in $D_1$, $x_{old}$ the image sample from the learned categories in $D_{99}$, $y$ the label “Yes”, $\bar{y}$ the label “No” and $\alpha$, $\beta$ are parameters used to balance the different contributions of the losses. Hyper-parameters in these experiments were set to $\alpha = 0.5$, $\beta = 0.001$, $\tau = 0.01$, and the learning rate $ lr = 0.01$. 

\subsection{Data expansion and translation module}
Building on the leave-one-out training procedure described above, this section details the data expansion process and translation module training that enable concept transfer between teacher and student networks. The datasets used in learning-by-communication experiment was CIFAR-100. The teacher Net was trained with dataset $D$ of all 100 categories, while the student Net was trained with $D_{99}$ containing 99 categories. An additional translation module was then trained to map the concept from teacher Net to student Net. Firstly, according to the procedure for training CATS Net, the teacher Net generated one concept for each category in $D$ ($D = D_{99} \cup D_1$), and the student Net generated one concept for each category in $D_{99}$. To generate enough samples for training this map, the teacher concept dataset was extended to 97 concept vectors for each category by concept vector expansion described above. Specifically, after the initial training of the teacher Net, the network parameters were fixed. Then 96 additional different concept vectors for each category were obtained through the training procedure described in the “Leave-one-out training and concept vector expansion” section. 

The translation module used was a multiple-layer perceptron, with ten hidden layers containing 500 neurons each. The ReLU activation function and the mean squared error (MSE) loss function were used. During the training, the dropout probability of all hidden layers was set to 0.3. The translation module was trained, for $D_{99}$, to map the 97 concept for each category from the teacher Net to the corresponding 1 concept for each category from the student Net. The learning rate was set to 0.0001 and decayed by the factor of 0.5 for every ten epochs. The Adam \cite{kingma_adam_2015} algorithm was used. The training lasted for 200 epochs to ensure convergence. The experiment was repeated 100 rounds, with a different class chosen as $D_1$ in each round.

\textbf{Semantic detail preservation analysis.} To assess whether the translation module preserves semantic details during concept transfer, we conducted layer-wise representational analysis across all 100 teacher-student pairs. For each translation module, we extracted feature vectors from the input layer, all 11 ReLU hidden layers, and the output layer when processing the teacher's 100 concept vectors. These 13-layer feature representations were analyzed using RDM correlation analysis for quantitative assessment of information preservation. For RDM analysis, we computed pairwise Euclidean distances between all concept representations within each layer, then calculated Spearman rank correlations between layer-wise RDMs. Statistical significance was assessed using two-tailed t-tests across the 100 translation modules. This analysis revealed systematic preservation of semantic relationships throughout the translation process, with gradual but controlled information compression across layers.

\subsection{Word2Vec as concept}
In these experiments, CATS Net was trained using the category-name word vectors as the predefined concept vector, which were provided by the fastText library \cite{mikolov_advances_2018}. We used the pretrained 300-dimensional English word vectors (cc.en.300.bin) and reduced them to 20 dimensions using fastText's built-in \texttt{reduce\_model()} function, which employs Principal Component Analysis (PCA) for dimensionality reduction. This 20-dimensional representation was chosen to match the dimensionality of our learned concept vectors, enabling direct comparison between the two concept spaces. The dataset was divided into two parts, $D_{99}$ and $D_1$, in the same way as in the leave-one-out concept abstraction experiment. CATS Net was directly trained by class label names, represented by their 20-dimensional Word2Vec embeddings, with images belonging to $D_{99}$. Then it was tested with the untrained class name corresponding to $D_1$ to identify the images. Experiments were also repeated 100 rounds with each category chosen as $D_1$.

\subsection{THINGS SPOSE49 and Binder65 as concept}
First, we identified 334 shared concepts between the ImageNet-1k and THINGS datasets. Both the ImageNet-1k and THINGS datasets provide category labels with unique synset ID in WordNet \cite{fellbaum_wordnet_1998}. By matching these IDs, we extracted 334 shared concepts. Feature vectors for these concepts were then extracted from the SPOSE49 model provided by Hebart et al. \cite{hebart_revealing_2020}. For Binder65, feature vectors for each object name were computed as Pearson’s correlation coefficients between the object name embeddings and the Binder 65 dimension name embeddings in the Word2Vec embedding space \cite{grand_semantic_2022}. Two concepts could not be represented in the Binder65 feature space, resulting in a final set of 332 concepts for subsequent analyses. RDMs were constructed using pairwise Pearson's distance (that is, 1 - Pearson's correlation coefficient) between feature vectors. 

For analyses focusing on specific Binder65 subdomains, RDMs were computed using the corresponding subset of Binder65 dimensions. The "cognition" domain was excluded from these analyses due to its single-dimensional structure, which precluded the calculation of meaningful dissimilarity matrices required for our analytical approach.

For the WT95 dataset, we identified 89 shared concepts between the WT95 stimulus set and the THINGS dataset. All subsequent analyses on this dataset were conducted using these 89 concepts.

\subsection{fMRI dataset} 
\textbf{Participants.} Twenty-nine participants ($19$ females; median age, $20$ years; range, $18$-$32$ years) were recruited in our study and were scanned in a conditional-rich event-related fMRI experiment. All participants were right-handed, native Mandarin speakers with normal or corrected-to-normal vision and had no history of neurological or language disorders. All protocols and procedures of the current study were approved by the State Key Laboratory of Cognitive Neuroscience and Learning at Beijing Normal University (ICBIR\_A\_0040\_008). Prior to participation, all participants provided written informed consent. The study was conducted in accordance with the Declaration of Helsinki and adhered to all relevant ethical guidelines.

\textbf{Stimulus and procedures.} Ninety-five objects were chosen, including $3$ common domains ($32$ animals, $35$ small manipulable artifacts, and $28$ large nonmanipulable artifacts). Each object was presented as a $400\times400$ pixels colored image displaying a representative exemplar against a white background ($10.55$\textdegree $\times$ $10.55$\textdegree of visual angle). The stimulus described above will hereafter be referred to as the WT95 object image dataset. All the participants were asked to name each displayed picture using oral language. The whole experiment included $6$ runs, with each item repeated for $6$ times across the experiment. Each run ($8$ min $45$ s) consisted of $95$ trials, with each item presented once per run. The trial structure consisted of a $0.5$ s fixation, followed by a $0.8$ s stimulus presentation and an inter-trial interval (ITI) ranging from $2.7$ s to $14.7$ s. The order of stimuli and ITI durations were randomized using the optseq2 optimization algorithm (\url{http://surfer.nmr.mgh.harvard.edu/optseq/}) (Dale, 1999). Each run began and ended with a $10$ s fixation period.

\textbf{Image acquisition.} Functional and anatomical MRI images were collected at the MRI center, Beijing Normal University using a $3$ Tesla Siemens Trio Tim Scanner. A high-resolution 3D structural image was collected with a 3D magnetisation prepared-rapid gradient echo (3D-MPRAGE) sequence in the sagittal plane ($144$ slices, TR = $2530$ ms, TE = $3.39$ ms, flip angle = $7$°, matrix size = $256\times256$, voxel size = $1.33\times 1\times1.33$ mm). Functional images were acquired with an echo-planar imaging (EPI) sequence (33 axial slices, TR = $2000$ ms, TE = $30$ ms, flip angle = $90$\textdegree, matrix size = $64\times64$, voxel size = $3\times3\times3.5$ mm with a gap of $0.7$ mm).

\subsection{WT95 RDM of CATS Net model}
We have previously trained 30 different CATS Nets on ImageNet-1k and these models possess distinct conceptual spaces. Based on these spaces, we abstracted the concepts from WT95 object image dataset to form the WT95 RDMs for each model. Specifically, for each CATS Net, we retained all the network parameter modules (TS module and CA module), discarded the concept set, and then allocated 95 random initial points in 20-dimension space as 95 concept vectors. Subsequently, with the network parameters fixed, we updated only the concept vectors through BP algorithm until convergence was achieved. The dissimilarity between each pair of concepts was then calculated as 1 – Pearson’s correlation coefficient to generate the $95\times95$ RDM. 

In order to obtain a more accurate estimation of the concept space of the model through the RDM, we repeated the above concept formation process 100 times for one model. Then, we averaged the RDMs of these 100 sets of concepts to represent the RDM of that model.

\subsection{Preprocessing for Task-fMRI data}
The functional images were preprocessed and analyzed using Statistical Parametric Mapping (SPM12; \url{http://www.fil.ion.ucl.ac.uk/spm}). For each participant, the first $5$ volumes of each run were discarded for signal equilibrium. Then the remaining images were corrected for time slicing and head motion and then spatially normalized to Montreal Neurological Institute (MNI) space via unified segmentation (resampling into $3\times3\times3$ mm$^3$ voxel size). Three subjects were excluded from the data analyses due to the successive head motions ($>3$ mm/$3$\textdegree). For the functional images of each participant, the object-relevant beta weights were obtained using general linear model (GLM). The GLM contained onset regressor for each of $95$ items, $6$ regressors of no interest corresponding to the $6$ head motion parameters, and a constant regressor for each run. Each item-relevant regressor was convolved with a canonical HRF, and a high-pass filter cutoff was set as $128$s. The resulting t-maps for each item versus baseline were used to create neural RDMs.

\subsection{ROI definition}
The VOTC mask was defined as regions showing stronger activation to all pictures relative to the rest in the fMRI dataset (FDR $q < 0.05$) within the cerebral mask combining the posterior and temporooccipital divisions of inferior temporal gyrus (15\#, 16\#), the inferior division of lateral occipital cortex (23\#), the posterior division of parahippocampal gyrus (35\#), the lingual gyrus (36\#), the posterior division of temporal fusiform cortex (38\#), the temporal occipital fusiform cortex (39\#), the occipital fusiform gyrus (40\#), the supracalcarine cortex (47\#) and the occipital pole (48\#) in the Harvard-Oxford Atlas (probability $> 0.2$).

\subsection{Representation similarity analysis}
\textbf{For the ROI-level analysis,} activity patterns for each item within each ROI were extracted from whole-brain \textit{t}-statistic images. Neural RDMs were then generated based on the Pearson distance between activation patterns for each object pair. Model fitting was quantified by computing the partial Spearman’s rank correlation (Spearman’s $\rho$) between the neural RDMs and model RDMs, controlling for RDMs derived from the feature extraction layer. For the analysis of the CA module specifically, this partialling out procedure was not applied. The resulting correlation coefficients underwent Fisher-\textit{z} transformation and were averaged at the subject level. At the model-group level, one-sample \textit{t}-tests were performed on the subject-level mean correlation coefficients ($\rho$ values) to determine significant differences from zero. For comparative analyses between model groups within VOTC, two-sample \textit{t}-tests were employed to evaluate differences in subject-level mean $\rho$ values. Additionally, paired \textit{t}-tests were utilized to assess statistical differences between ROIs, enabling direct comparison of regional effects across the predefined functional and anatomical boundaries.

\textbf{For the whole-brain analysis,} a searchlight approach was implemented wherein multivariate activation patterns within a sphere (radius = 10 mm) centered on each voxel were extracted to compute Pearson-based neural RDMs. For each searchlight position, the Spearman’s rank correlation (Spearman’s $\rho$) between the neural RDM and model-derived RDMs was computed, partialling out the effects from the sensory input layer. For the analysis of the CA module specifically, this partialling out procedure was not applied. This procedure generated correlation maps for each participant by iteratively moving the searchlight center throughout the whole brain. The resulting correlation maps underwent Fisher-\textit{z} transformation and were spatially smoothed using a 6 mm full-width at half-maximum (FWHM) Gaussian kernel. These processed maps were then averaged across subjects to produce group-level representation. Statistical significance was assessed through model-group-level analysis using one-sample \textit{t}-tests against zero to identify brain regions showing significant correlations with the theoretical models.

\subsection{Noise ceiling estimation}
Our primary effect size for each model instance \textit{i} is computed by correlating the instance’s RDM with each participant’s RDM within a ROI using Spearman’s $\rho$, applying a Fisher-\textit{z} transform, and averaging across participants. So, to provide an upper bound that is commensurate with this statistic, we estimate a noise ceiling (NC) in the \textit{z}-domain that jointly reflects measurement reliability on the participant side and stochastic variability on the model-instance side.

\textbf{For participant-wise reliability} $rel_s$, we first estimate the group-mean RDM reliability $rel_{group}$ via participant split-half half-sample means correlated across many random 50/50 splits, Fisher-\textit{z} averaged and Spearman-Brown corrected), then relate each participant to the leave-one-out group mean, $r(X_s, \overline{X_{-s}})$, and obtain:
\[
rel_s \ge \frac{r^2(X_s, \overline{X_{-s}})}{rel_{group}}
\]

\textbf{For single-instance model reliability} $rel_{model}$, we estimate the reliability of a single model instance using a leave-one-out approximation: for each instance \textit{i}, we compute the correlation between $M_i$, and the mean RDM of the remaining $M-1$ instances, $r_i = \rho(M_i, \overline{M_{-i}})$; we then average $r_i$ in the Fisher-\textit{z} domain and back-transform to obtain $rel_{model}$.

Finally, the expected correlation between a single participant and a single instance is bounded by $\sqrt{rel_{s}rel_{model}}$. Because our effect averages Fisher-\textit{z} values across participants, we aggregate the bound in the same domain:
\[
NC_z = \frac{1}{S}\sum_{s=1}^{S}\atanh(\sqrt{rel_{s}rel_{model}})
\]
 
\subsection{Brain visualization}
The brain results were projected onto the MNI brain surface for visualization using BrainNet Viewer \cite{xia_brainnet_2013} (version 1.7; \url{https://www.nitrc.org/projects/bnv/}; RRID: SCR\_009446) with the default 'interpolated' mapping algorithm, unless stated explicitly otherwise.

\subsection{Model RDM clustering}
K-means clustering was performed using the \textit{kmeans} function in Matlab R2021a with default parameters ($k = 2$).

\subsection{Statistics \& Reproducibility}
\textbf{Sample size determination,} no statistical methods were used to pre-determine sample sizes, but our sample size (N=26) is similar to those reported in previous publications investigating semantic representations (for instance, \cite{fairhall_brain_2013,devereux_representational_2013,tyler_objects_2013}). 

\textbf{Data exclusion,} out of the 29 subjects recruited, data from 3 subjects were excluded from the final analyses because of excessive head motion ($>3$ mm or 3°). No other data were excluded.

\textbf{Randomization and blinding,} the experiments were not randomized as there were no group allocations involved in this study. The Investigators were not blinded to allocation during experiments.

\textbf{Assumptions of the statistical tests,} data distribution was assumed to be normal but this was not formally tested. The data distributions and individual data points were all plotted.

\section{Data Availability}
Source data are provided with this paper. The fMRI data that support the findings of this study have been deposited in the Open Science Framework (OSF) at \url{https://osf.io/5y8p6/overview}. The embeddings of SPOSE49 model \cite{{hebart_revealing_2020}} are available via OSF at \url{https://osf.io/f5rn6/files/8yjh5}. Additionally, the anchor word embeddings used for the Binder65 model \cite{binder_toward_2016} can be accessed at \url{https://www.neuro.mcw.edu/index.php/resources/brain-based-semantic-representations/}. We only used the ImageNet-1k training part and validation part for CATS Net training and testing in this work. Website: \url{https://image-net.org/index.php}. Website for CIFAR 100 dataset \cite{krizhevsky_learning_2009}: \url{https://www.cs.toronto.edu/~kriz/cifar.html}.

\section{Code availability}
The code source of all results shown in this Article is available via GitHub and Zenodo at \url{https://doi.org/10.5281/zenodo.18136642}.

\section{Acknowledgements}

This research was supported by grants from the CAS Project for Young Scientists in Basic Research (grant No. YSBR-041 to Y.C.); the National Natural Science Foundation of China (grant Nos. 32595490, 32595491 to Y.B.); Strategic Priority Research Program of the Chinese Academy of Sciences (CAS) (grant No. XDB1010302 to S.Y.); the International Partnership Program of Chinese Academy of Sciences (grant No. 104GJHZ2025032FN to Y.C.); the STI2030-Major Project 2021ZD0204100 (grant No. 2021ZD0204104 to Y.B.); the National Natural Science Foundation of China (grant Nos. 31925020 and 82021004 to Y.B.). The funders had no role in the study design, data collection and analysis, decision to publish or preparation of the manuscript. We thank Danko Nikolić, Frederic Alexandre, Jinpeng Zhang, Guoqing Ma, Xiaosha Wang and Huichao Yang for their valuable comments on earlier drafts of the manuscript.

\section{Author Contributions Statement}

L.G. and H.C. conceived the study under supervision of Y.C., Y.B. and S.Y.; L.G., H.C. and Y.C. designed the experiment; L.G. and Y.C. implemented and conducted the experiments on CATS Net models; H.C. and L.G analyzed the models and fMRI data and plotted the results; L.G., H.C. and Y.C. wrote the initial draft; all authors reviewed and edited the Article.

\section{Competing Interests Statement}

The authors declare no competing interests.

\newpage


\section{Figure Legends}

\begin{figure}[htbp]
\centering
\includegraphics[width=0.8\textwidth]{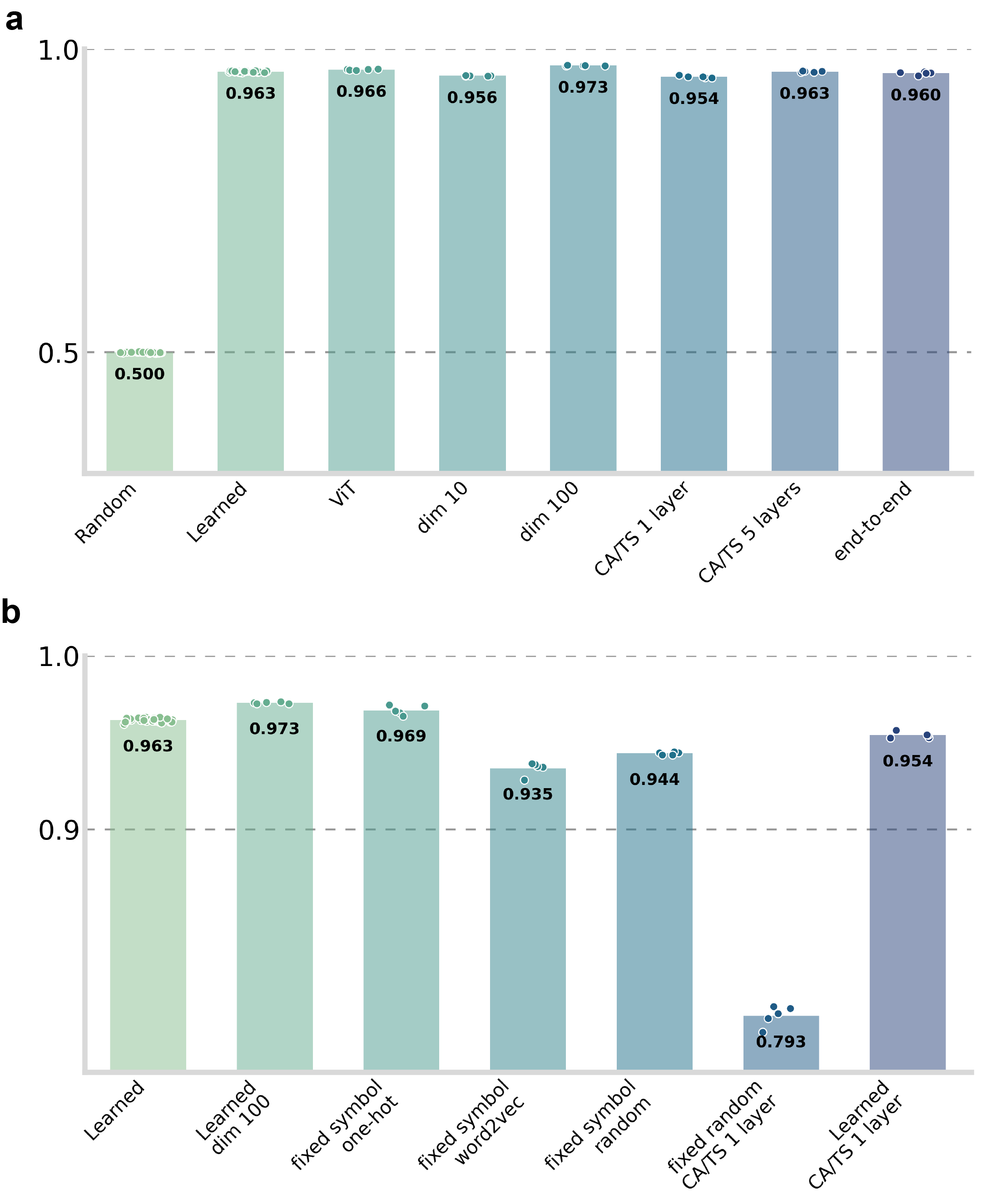}
\suppcaption{\textbf{$\vert$ Ablation study on CATS Net. a,} Ablation studies and hyperparameter explorations on backbone, concept size, number layers of CA/TS module and training strategy. The left most 2 bars was adopted from Fig~\ref{fig2}a, while the others represents the average of mean accuracy across 5 independently initialized models after training, and each point represents the corresponding mean accuracy cross all categories (from the $3^{rd}$ bar to the right most one: using ViT as backbone, setting concept size to 10, setting concept size to 100, setting the layer number of CA/TS module to 1, setting the layer number of CA/TS module to 5, end-to-end training of concept vectors together with CA/TS module). \textbf{b,} Ablation studies on concept space construction. The left most one bar was adopted from Fig~\ref{fig2}a, while the others represents the average of mean accuracy across 5 independently initialized models after training, and each point represents the corresponding mean accuracy cross all categories (from the $2^{nd}$ bar to the right most one: setting concept size to 100, setting concept size to 1000 using fixed one-hot vectors, setting concept to fixed Word2Vec vectors projected to 20 dimensions, setting concept to fixed 20-dimension random vectors, setting concept to fixed 20-dimension random vectors with 1 CA/TS layer, setting 20-dimension concept vectors to be learnable with 1 CA/TS layer).}
\label{supp_fig1}
\end{figure}

\begin{figure}[htbp]
\centering
\includegraphics[width=0.9\textwidth]{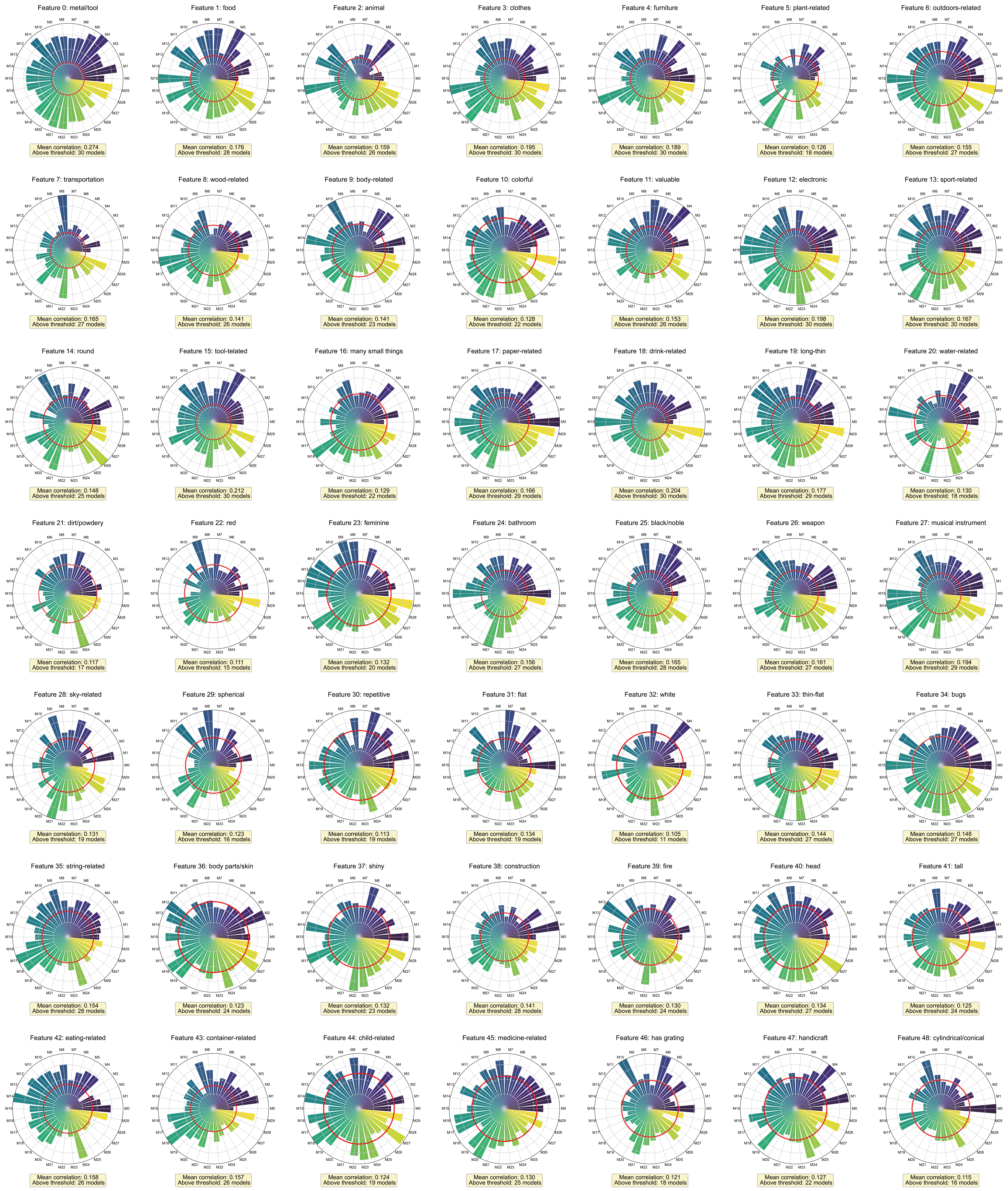}
\suppcaption{\textbf{$\vert$ Maximum correlations between CATS instances and all SPOSE49 dimensions. } Each bar represents the maximum Pearson correlation between the 20 concept dimensions of a given CATS instance and each SPOSE49 dimension (dimension labels shown around the perimeter; dimension names from Hebart et al. \cite{hebart_revealing_2020}). The red circle indicates the significance threshold ($r = 0.107$, two-tailed $p < 0.05$, df = 330).}
\label{supp_fig2}
\end{figure}

\begin{figure}[htbp]
\centering
\includegraphics[width=0.9\textwidth]{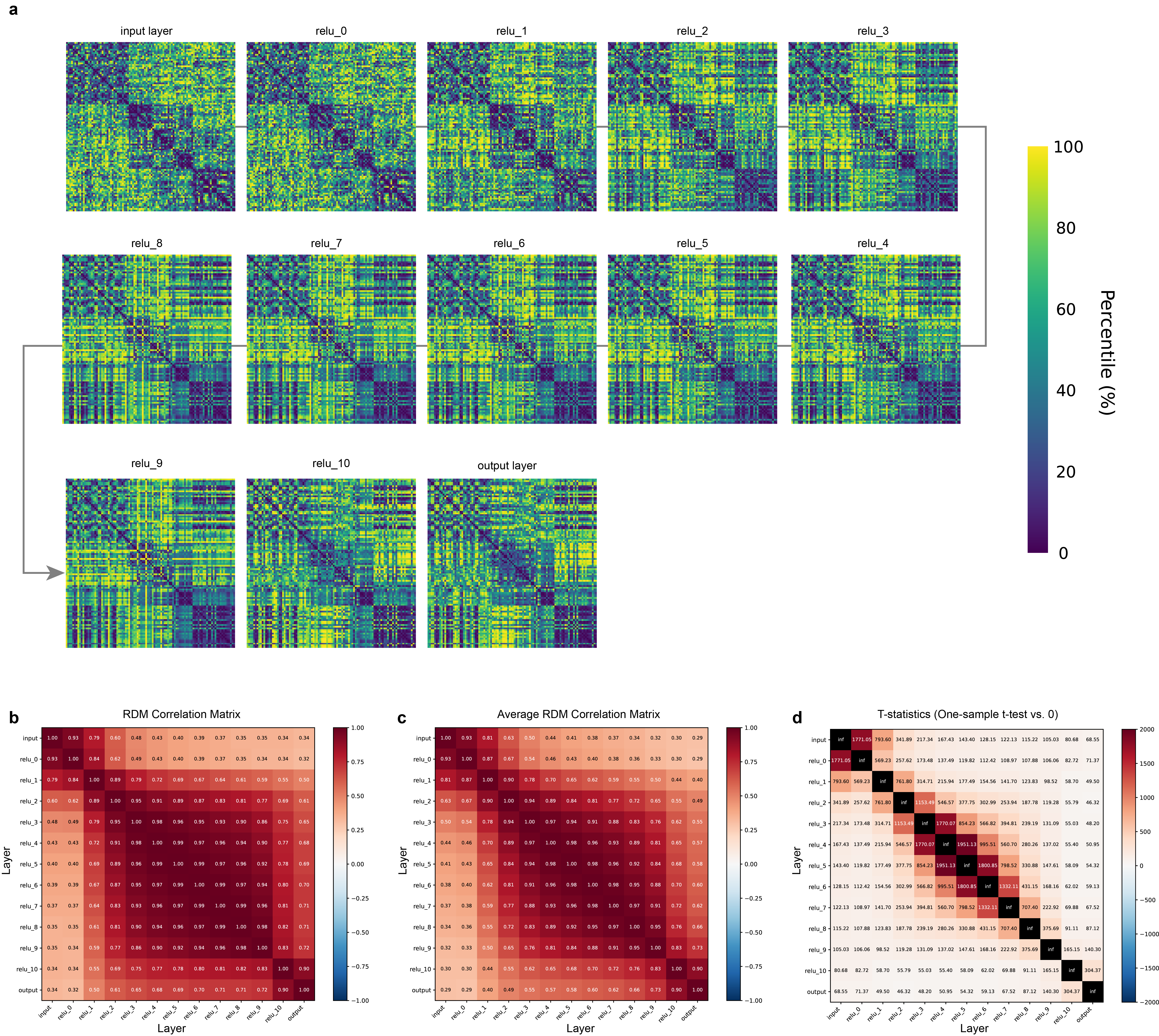}
\suppcaption{\textbf{$\vert$ Translation Module Analyses. a,} For this translation module ("apple" category was withheld from the student Net's training), given all 100 teacher concept vectors as input, we recorded the layer-wise activation and conducted layer-wise RDM (Pearson's correlation). \textbf{b,} The layer-wise RDM Spearman's correlation similarity matrix based on (a). \textbf{c,} The average layer-wise RDM Spearman's correlation similarity across all 100 translation modules. \textbf{d,} One-sample t-test of each value at translation module group level.}
\label{supp_fig3}
\end{figure}

\begin{figure}[htbp]
\centering
\includegraphics[width=0.9\textwidth]{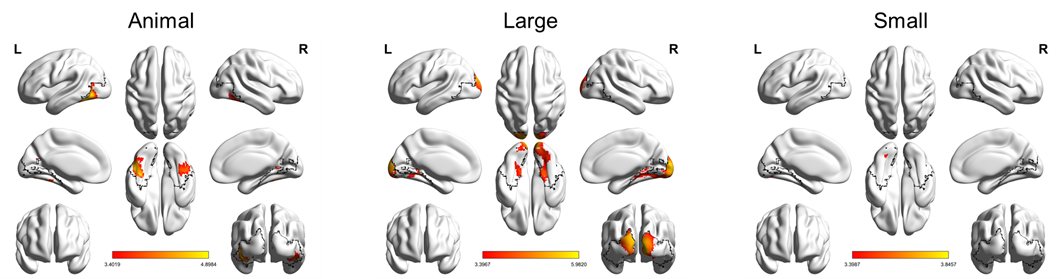}
\suppcaption{\textbf{$\vert$ Searchlight RSA results within the VOTC mask for three common semantic categories (animals, large non-manipulable objects, small manipulable objects).} The maps show \textit{t}-values reflecting the model-instance-level correspondence between the CATS concept layer representations (n=30) and brain activity of 26 subjects. Results are thresholded at voxel-level $p < 0.001$ (one-tailed) and cluster-level family-wise error (FWE) corrected $p < 0.05$. The color scale represents the $t$-statistic values.}
\label{supp_fig4}
\end{figure}

\clearpage



\end{document}